\documentclass[lettersize,journal]{IEEEtran}
\usepackage{amsmath,amssymb,amsfonts}
\usepackage{algorithmic}
\usepackage{algorithm}
\usepackage{array}
\usepackage{textcomp}
\usepackage{stfloats}
\usepackage{url}
\usepackage{verbatim}
\usepackage{graphicx}
\usepackage{cite}
\usepackage{bm}
\usepackage{xcolor}
\usepackage{hyperref}
\usepackage{multirow}
\usepackage{booktabs}
\usepackage{siunitx}
\usepackage{array}
\usepackage{tikz}
\usepackage{subfigure}

\begin{document}

\title{Radio-Based Passive Target Tracking by a Mobile Receiver with Unknown Transmitter Position}

\author{Ke~Xu, 
        Rui~Zhang, 
        and~He~(Henry)~Chen

\thanks{Ke Xu is with the Department of Information Engineering, The Chinese University of Hong Kong, Hong Kong SAR, China (e-mail: xk020@ie.cuhk.edu.hk).

Rui Zhang was with the Department of Information Engineering, The Chinese University of Hong Kong, Hong Kong SAR, China. He is now with the College of Electronics and Information Engineering, Shenzhen University, China (e-mail: zhrrygh@szu.edu.cn).

He (Henry) Chen is with the Department of Information Engineering, and the Shun Hing Institute of Advanced Engineering, The Chinese University of Hong Kong, Hong Kong SAR, China (e-mail: he.chen@ie.cuhk.edu.hk).}
}

\maketitle

\begin{abstract}

In this paper, we propose a radio-based passive target tracking algorithm using multipath measurements, including the angle of arrival and relative distance. We focus on a scenario in which a mobile receiver continuously receives radio signals from 
a transmitter located at an unknown position. The receiver utilizes multipath measurements extracted from the received signal to jointly localize the transmitter and the scatterers over time, with scatterers comprising a moving target and stationary objects that can reflect signals within the environment. We develop a comprehensive probabilistic model for the target tracking problem, incorporating the localization of the transmitter and scatterers, the identification of false alarms and missed detections in the measurements, and the association between scatterers and measurements. We employ a belief propagation approach to compute the posterior distributions of the positions of the scatterers and the transmitter. Additionally, we introduce a particle implementation for the belief propagation method. 
Simulation results demonstrate that our proposed algorithm outperforms existing benchmark methods in terms of target tracking accuracy.

\end{abstract}


\section{Introduction}\label{introduction}
\IEEEPARstart{T}{arget}
tracking is a critical technology with diverse applications, such as indoor localization, secure surveillance, and autonomous driving. It has been extensively studied for several decades due to its relevance in both civilian and military contexts \cite{mazor1998interacting,blackman2004multiple,meyer2018message}. The goal in a target tracking problem is to continuously estimate the position of one or multiple targets over time using sensor measurements. Some target tracking methods, such as those using cameras or active sensors, can directly measure target information like distance and angle with respect to the sensor\cite{morbidi2012active,zhou2011multirobot}. However, these systems heavily rely on favorable lighting conditions or the active cooperation of the target, such as emitting radio-frequency (RF) signals. 

In the wireless research area, a scenario of interest is when the target does not carry any RF devices or actively participate in the tracking process. This is commonly studied in the radar community \cite{niu2012target,gorji2012multiple}, in which the RF signals emitted from a transmitter are reflected by the target and then received by the radar receiver, and the target position is determined from the received signal. Radar systems can be further categorized into two types: active radar and passive radar \cite{franken2022integrating,zhang2022convex,dai2022composed}.\footnote{It is worth mentioning that the term ``passive'' in this context is different from its meaning in ``passive target tracking,'' where the latter refers to a scenario in which the target does not carry an RF device.}  In active radar, the transmitter is part of the system and can emit signals with a known waveform. On the other hand, in passive radar, the transmitter is not part of the system; instead, existing infrastructure in the environment, such as digital television, WiFi, and cellular networks, serves as the transmitter (often referred to as the illuminator) to illuminate a surveillance area \cite{berger2010signal,zhang2017multistatic}. This approach not only reduces the energy cost but also enhances anonymity and avoids interference with surrounding wireless systems. However, since the transmitter is non-cooperative, synchronization between the transmitter and the receiver is not possible.

Many existing works in passive radar-like systems are based on the assumption that the transmitter position is known \cite{malanowski2012two,hack2014detection,yi2015mimo,guruacharya2020map,zaimbashi2024integrated,meyer2017scalable}. This assumption may not hold in practice, especially when the system operates in a new environment. 
Several studies have attempted to address the problem of target localization and tracking with unknown transmitter positions. In \cite{zhang2019multistatic} and \cite{zheng2020accurate}, the authors have investigated the localization of a stationary target with an unknown transmitter position using the propagation distance of the direct and reflected signals, assuming perfect synchronization between the transmitter and multiple receivers. In \cite{zhang2022localization} and \cite{Panwar2024multistatic}, the target localization scenario is extended to the case where synchronization between the transmitter and receivers, as well as among receivers, is unavailable.  In \cite{wu2021efficient}, a multiple-input multiple-output (MIMO) system with unknown transmitter positions is considered, also addressing the lack of synchronization between transmitters and receivers. In \cite{wang2023multistatic}, the target localization problem in the absence of the transmitter position and synchronization is addressed aided by a calibration object with a known position.  Furthermore, in \cite{zhang2020multistatic,zhang2021multistatic,wang2021elliptic,pei2023reducing}, scenarios are examined where at least one of the target or the transmitter is moving. In \cite{li2022multistatic}, the target localization problem is investigated when both the transmitter position and the signal propagation speed are unknown. In all the above cases, multiple receivers deployed at different positions are required for localization based on distance-related measurements (and frequency-related measurements in the moving scenarios). In \cite{zhang2023localization}, localization of transmitters and scatterers by a single receiver is investigated based on time difference of arrival (TDOA) and angle of arrival (AOA), though it requires intentional transmitters or scatterers with known positions to assist in localization. In \cite{yang2017adaptive} and \cite{yang2018target}, target tracking problems with at least two transmitters are explored, where the signal from one transmitter needs to be reflected by the other. In \cite{michaelis2015bistatic}, a single receiver is used to jointly localize the transmitter and multiple landmarks that reflect signals based on relative delay, AOA, and Doppler frequency. All of the aforementioned works, however, do not take into account false alarms and missed detections in the signal measurements, which means that a multipath measurement may not correspond to a true scatterer in the environment, and a true scatterer may also fail to generate a measurement. Neglecting these issues can result in significant localization errors.

In this paper, we consider a passive target tracking problem with an unknown transmitter position, in which a mobile receiver continuously collects multipath measurements from received signals to estimate the target and transmitter positions over time. The mobility of the receiver enables the collection of measurements from different locations over time, allowing the system to determine both the transmitter and target positions effectively without requiring multiple receivers. Since the transmitter is non-cooperative, synchronization between the transmitter and the receiver is not feasible. Consequently, only the AOA of the direct and scattered paths, as well as the relative delay between them, can be measured. Compared to previous literature, our work incorporates the multipath effect caused by stationary objects in the environment. The receiver extracts not only relative distance and AOA measurements associated with the target but also those associated with the stationary objects, leading to an association problem between the scatterers (including the target and all stationary objects) and measurements. In addition, we account for the false alarms and missed detections in the measurements. We develop a probabilistic model to capture all these factors. Since the model is highly complicated, direct computation of the posterior distributions of target positions is unfeasible. Therefore, we propose a belief propagation-based algorithm, along with a particle implementation, to efficiently solve the problem. The main contributions of the paper can be summarized as follows:

\begin{itemize}
\item
We formulate a probabilistic model for the passive target tracking problem, taking into account the estimation of both the transmitter position and all scatterer positions over time, the identification of false alarms and missed detections in multipath measurements, and the association between the measurements and scatterers. Once all scatterer positions are obtained, the one exhibiting significant movement can be identified as the target position.

\item We represent the probabilistic model using a factor graph and develop a belief propagation-based algorithm to solve the target tracking problem. By passing messages on the graph, we can compute the posterior distributions of the transmitter and scatterer positions. Additionally, we introduce a low-complexity particle implementation to approximate the message computation.

\item We evaluate the performance of the proposed algorithm through simulations and compare it with three benchmarks and two simplified variants. Simulation results demonstrate that our proposed algorithm achieves superior performance in terms of target tracking accuracy compared to the benchmarks.

\end{itemize}

It is worth noting that our proposed belief propagation-based target tracking algorithm is inspired by the existing algorithm for simultaneous localization and mapping (SLAM)\cite{leitinger2019belief}, in which joint estimation of the positions of a moving receiver and multiple static physical or virtual transmitters is studied.  
We also note that a related problem to the one addressed in this paper has been studied in \cite{guo2017passive}. The algorithm employs an approximated linear model and primarily relies on the probabilistic data association (PDA) filter and joint PDA (JPDA) filter for associating targets with measurements and updating target positions, with additional heuristics applied to address the initialization and maintenance of target tracks. In contrast, our algorithm establishes a fully probabilistic framework that incorporates these considerations and retains model nonlinearity in the messages passed on the factor graph.

The rest of the paper is organized as follows. In Section \ref{signal model}, we introduce the system model and provide a probabilistic formulation for the passive target tracking problem. In Section \ref{bp algorithm}, we elaborate on the belief propagation approach to localize the transmitter and scatterer positions.  In Section \ref{particle implementation}, we introduce a particle implementation for approximating the message computations. In Section \ref{simulation}, simulations are conducted to validate the effectiveness of the proposed algorithm. Finally, we conclude the paper in Section \ref{conclusion}.

\textit{Notations:} Column vectors and matrices are represented by boldface lowercase letters and boldface uppercase letters, respectively. Sets are represented by calligraphic uppercase letters. The $n$-th entry of a vector $\mathbf{a}$ is denoted by $a_n$. $(\cdot)^\mathrm{T}$ and $\lVert\cdot\rVert_2$ denote the transpose and the Euclidean norm of a column vector, respectively. $\lvert\cdot\rvert$ denotes the cardinality of a set. $\mathrm{tr}(\cdot)$ denotes the trace of a matrix. 
$\arccos(\cdot)$ is the arccosine function of a real number between $-1$ and $1$. The probability density function (pdf) of a continuous random variable $x$ is denoted by $f(x)$, and the probability mass function (pmf) of a discrete random variable $x$ is denoted by $p(x)$. $\mathbf{I}_N$ is an identity matrix of size $N$. $\mathcal{N}(\mathbf{x};\bm{\mu},\mathbf{\Sigma})$ represents the pdf of a Gaussian distributed random vector $\mathbf{x}$ with mean $\bm{\mu}$ and covariance matrix $\mathbf{\Sigma}$.

\section{System Model and Problem Formulation}\label{signal model}
In this section, we introduce the system model for passive target sensing, and provide a probabilistic characterization of the problem. 
\subsection{System Model}\label{system model}

We consider a passive target tracking model as follows. A stationary transmitter is located at $\mathbf{x}_0$ and continuously emits radio signals to the surrounding wireless devices. The transmitted signal experiences a multipath channel and is then received by a mobile receiver. At time $n$, let $L_n$ denote the number of scatterers in the environment, including a moving target and other stationary objects. 
The receiver obtains an estimate of $L_n$ and extracts the multipath parameters from the received signal. We assume that the receiver is equipped with multiple antennas and therefore can estimate the AOA of both the direct path and the scattered paths, as well as the relative delay between the scattered paths and the direct path. For $\ell\in\{1,\cdots,L_n\}$, let $\mathbf{x}_{n,\ell}$ denote the position of the $\ell$-th scatterer,\footnote{It is worth noting that we use a unified indexing system and do not distinguish the moving target from other stationary objects. In addition, the passive target tracking algorithm proposed later can accommodate multi-target scenarios. However, we focus on a single target for clarity in both the description and simulation illustrations.} and let $\mathbf{x}_n$ denote the position of the receiver. The relative distance (directly proportional to the relative delay) and the AOA of the $\ell$-th scattered path can be written as
\begin{equation}\label{relative distance}
d_{n,\ell} = \lVert\mathbf{x}_{n,\ell}-\mathbf{x}_0\rVert_2+\lVert\mathbf{x}_n-\mathbf{x}_{n,\ell}\rVert_2-\lVert\mathbf{x}_0-\mathbf{x}_n\rVert_2,
\end{equation}
and
\begin{equation}\label{AOA}
\theta_{n,\ell} = \arccos\left( \frac{(\mathbf{x}_{n,\ell}-\mathbf{x}_n)^\mathrm{T}\mathbf{q}_n}{\lVert\mathbf{x}_{n,\ell}-\mathbf{x}_n\rVert_2}   \right),
\end{equation}
respectively, where $\mathbf{q}_n$ is a unit vector representing the orientation of the receiver. In (\ref{AOA}), we define the AOA as the angle between the receiver orientation and the direction pointing from the receiver to the scatterer. We note that the AOA of the direct path, denoted by $\theta_{n,0}$, can be expressed similarly to (\ref{AOA}) by replacing $\mathbf{x}_{n,\ell}$ with $\mathbf{x}_{0}$. In this paper, we assume that the position and the orientation of the receiver are known \textit{a priori}, which, for example, can be obtained in practice using a high-precision odometry sensor.

We assume that the relative distance and AOA extracted from received signals have estimation errors. These errors include incorrect estimations of the number of scatterers as well as inaccuracies in specific parameter values. Let $M_n$ denote the estimated number of scatterers at time $n$, which can be larger than, equal to, or smaller than $L_n$. This is because a false alarm measurement can be generated, which does not correspond to any scatterer in the environment, and a missed detection can also occur, resulting in a true scatterer not being detected and thus no measurement being generated.
We denote by $\mathbf{z}_n = [z_{n,0},\mathbf{z}_{n,1}^\mathrm{T},\cdots,\mathbf{z}_{n,M_n}^\mathrm{T}]^\mathrm{T}$ the collection of all measurements obtained at time $n$, where $z_{n,0}$ is the measured AOA of the direct path, and for $m\in\{1,\cdots,M_n\}$, $\mathbf{z}_{n,m}=[z_{d,n,m},z_{\theta,n,m}]^\mathrm{T}$ includes the measured relative distance and AOA of a true scattered path or a false alarm.\footnote{In practice, identification algorithms such as \cite{zhou2015wifi,zhang2019nothing,huang2020machine} can be used to detect the direct path. If no direct path is detected at certain time instants, the AOA of the direct path and the relative distances of scattered paths are not available; therefore, measurements from these moments can be discarded. The tracking algorithm resumes when the direct path is detected again.}

The task of the receiver is to use the multipath measurements to localize the scatterers and the transmitter, while simultaneously identifying false alarms and missed detections in the measurements and matching scatterers across consecutive time instants. Next, we will develop a probabilistic model to formally frame the problem.

\subsection{Probabilistic Formulation of Passive Target Tracking}\label{probabilistic formulation}

Although the transmitter remains stationary, the receiver updates its position estimate at each time instant. To account for this, we introduce a time index and use $\mathbf{x}_{n,0}$ to represent the transmitter position hereafter, and its transition pdf is denoted by $f(\mathbf{x}_{n,0}|\mathbf{x}_{n-1,0})$. 
To characterize the transitions of scatterer positions, we adopt the terminology similar to \cite{meyer2018message,leitinger2019belief}, referring to a previously detected scatterer as a legacy potential scatterer (PS), and a scatterer corresponding to a measurement at the current time instant as a new PS. At time $n$, the number of legacy PSs is denoted by $K_{n-1}$, and the number of new PSs is equal to the estimated number of scatterers $M_n$. The relationship between the numbers of legacy PSs at two consecutive time instants is given by $K_n=K_{n-1}+M_n$. In other words, once the new PSs at time $n$ have been detected, they will become legacy PSs at time $n+1$. For $k\in\{1,\cdots,K_{n-1}\}$, the position of the $k$-th legacy PS is denoted by $\underline{\mathbf{x}}_{n,k}$. In addition, a binary variable $\underline{r}_{n,k}\in\{0,1\}$ is used to indicate the existence of the legacy PS. Specifically, if $\underline{r}_{n,k}=1$, the legacy PS exists; otherwise, it does not. The non-existence of a legacy PS can be due to the disappearance of the scatterer in a dynamic environment. For example, this might happen if the target acting as a scatterer exits the observed area. We define the augmented state of the $k$-th PS as $\underline{\mathbf{y}}_{n,k}\triangleq[\underline{\mathbf{x}}_{n,k}^\mathrm{T},\underline{r}_{n,k}]^\mathrm{T}$. With similar notations, for $m\in\{1,\cdots,M_n\}$, we denote the position of the $m$-th new PS by $\overline{\mathbf{x}}_{n,m}$, and the existence indicator by $\overline{r}_{n,m}$. If $\overline{r}_{n,m}=1$, the $m$-th measurement is generated by a scatterer detected for the first time; otherwise, it is not.  
The augmented state of the $m$-th new PS is defined as $\overline{\mathbf{y}}_{n,m}\triangleq[\overline{\mathbf{x}}_{n,m}^\mathrm{T},\overline{r}_{n,m}]^\mathrm{T}$. Moreover, for $k\in\{1,\cdots,K_n\}$, we use $\mathbf{x}_{n,k}$ and $r_{n,k}$ to represent the position and the existence indicator of a general PS, respectively, and denote the augmented state by $\mathbf{y}_{n,k}\triangleq[\mathbf{x}_{n,k}^\mathrm{T},r_{n,k}]^\mathrm{T}$.
Let $f(\underline{\mathbf{y}}_{n,k}|\mathbf{y}_{n-1,k})=f(\underline{\mathbf{x}}_{n,k},\underline{r}_{n,k}|\mathbf{x}_{n-1,k},r_{n-1,k})$ be the transition pdf of the augmented state. If $r_{n-1,k}=0$, then we have
\begin{equation}\label{transition0}
f(\underline{\mathbf{x}}_{n,k},\underline{r}_{n,k}|\mathbf{x}_{n-1,k},0)=    
\begin{cases}
    f_\mathrm{D}(\underline{\mathbf{x}}_{n,k}),&\underline{r}_{n,k}=0,\\
    0,&\underline{r}_{n,k}=1,\\
\end{cases}
\end{equation}
where $f_\mathrm{D}(\underline{\mathbf{x}}_{n,k})$ is a dummy pdf that integrates to $1$. The above equation indicates that if a PS does not exist at time $n-1$, it cannot exist at time $n$. On the other hand, if $r_{n-1,k}=1$, i.e., the PS exists at time $n-1$, it will survive at the next time instant with probability $p_\mathrm{s}$. Therefore, we have
\begin{equation}\label{transition1}
f(\underline{\mathbf{x}}_{n,k},\underline{r}_{n,k}|\mathbf{x}_{n-1,k},1)=    
\begin{cases}
    (1-p_\mathrm{s})f_\mathrm{D}(\underline{\mathbf{x}}_{n,k}),&\underline{r}_{n,k}=0,\\
    p_\mathrm{s}f(\underline{\mathbf{x}}_{n,k}|\mathbf{x}_{n-1,k}),&\underline{r}_{n,k}=1,\\
\end{cases}
\end{equation}
where $p_\mathrm{s}$ is the survival probability of an existing PS, and $f(\underline{\mathbf{x}}_{n,k}|\mathbf{x}_{n-1,k})$ is the transition pdf of its position if it survives.

Since it is not known at each time instant which measurement is generated by which legacy PS, or whether a measurement is a false alarm, we need to perform association between the legacy PSs and the measurements. Following \cite{williams2010data,williams2014approximate,meyer2018message,leitinger2019belief}, we assume that each measurement is generated by at most one scatterer and that each scatterer can generate at most one measurement, and then define the \textit{scatterer-oriented} association vector as $\mathbf{a}_n=[a_{n,1},\cdots,a_{n,K_{n-1}}]^\mathrm{T}$, where $a_{n,k}=m\in\{1,\cdots,M_n\}$ if the $k$-th PS generates the $m$-th measurement $\mathbf{z}_{n,m}$, and $a_{n,k}=0$ if it does not generate a measurement. In addition, we define the \textit{measurement-oriented} association vector as $\mathbf{b}_n=[b_{n,1},\cdots,b_{n,M_n}]^\mathrm{T}$, where $b_{n,m}=k\in\{1,\cdots,K_{n-1}\}$ if the $m$-th measurement is generated by the $k$-th legacy PS, and $b_{n,m}=0$ if it is a false alarm. It is evident that $\mathbf{a}_n$ and $\mathbf{b}_n$ contain the same information because they can be determined from each other. However, as will be seen, this redundant formulation can help ``open'' or ``stretch'' certain factors in the probabilistic model, thereby making the target tracking algorithm more scalable.

To facilitate subsequent derivations, we denote $\underline{\mathbf{y}}_n\triangleq[\underline{\mathbf{y}}_{n,1}^\mathrm{T},\cdots,\underline{\mathbf{y}}_{n,K_{n-1}}^\mathrm{T}]^\mathrm{T}$, $\underline{\mathbf{x}}_n\triangleq[\underline{\mathbf{x}}_{n,1}^\mathrm{T},\cdots,\underline{\mathbf{x}}_{n,K_{n-1}}^\mathrm{T}]^\mathrm{T}$, $\underline{\mathbf{r}}_n\triangleq[\underline{r}_{n,1},\cdots,\underline{r}_{n,K_{n-1}}]^\mathrm{T}$, $\overline{\mathbf{y}}_n\triangleq[\overline{\mathbf{y}}_{n,1}^\mathrm{T},\cdots,\overline{\mathbf{y}}_{n,M_n}^\mathrm{T}]^\mathrm{T}$, $\overline{\mathbf{x}}_n\triangleq[\overline{\mathbf{x}}_{n,1}^\mathrm{T},\cdots,\overline{\mathbf{x}}_{n,M_n}^\mathrm{T}]^\mathrm{T}$, $\overline{\mathbf{r}}_n\triangleq[\overline{r}_{n,1},\cdots,\overline{r}_{n,M_n}]^\mathrm{T}$, 
$\mathbf{y}_n\triangleq[\underline{\mathbf{y}}_n^\mathrm{T},\overline{\mathbf{y}}_n^\mathrm{T}]^\mathrm{T}$, $\mathbf{x}_n\triangleq[\underline{\mathbf{x}}_n^\mathrm{T},\overline{\mathbf{x}}_n^\mathrm{T}]^\mathrm{T}$, $\mathbf{r}_n\triangleq[\underline{\mathbf{r}}_n^\mathrm{T},\overline{\mathbf{r}}_n^\mathrm{T}]^\mathrm{T}$, 
$\mathbf{x}_{1:n,0}\triangleq[\mathbf{x}_{1,0}^\mathrm{T},\cdots,\mathbf{x}_{n,0}^\mathrm{T}]^\mathrm{T}$, $\mathbf{y}_{1:n}\triangleq[\mathbf{y}_1^\mathrm{T},\cdots,\mathbf{y}_n^\mathrm{T}]^\mathrm{T}$, $\mathbf{a}_{1:n}\triangleq[\mathbf{a}_1^\mathrm{T},\cdots,\mathbf{a}_n^\mathrm{T}]^\mathrm{T}$, $\mathbf{z}_{1:n}\triangleq[\mathbf{z}_1^\mathrm{T},\cdots,\mathbf{z}_n^\mathrm{T}]^\mathrm{T}$, and $\mathbf{m}_{1:n}\triangleq[M_1,\cdots,M_n]^\mathrm{T}$. We note that $M_n$ is unknown and considered as a random variable before the measurement $\mathbf{z}_n$ is observed at each time instant. Once $\mathbf{z}_n$ is observed, $M_n$ becomes known and fixed. 
The primary objective of the passive target tracking problem is to detect the existence of each PS and estimate the positions of the existing scatterers at each time instant from a joint posterior pdf $f(\mathbf{x}_{1:n,0},\mathbf{y}_{1:n},\mathbf{a}_{1:n}|\mathbf{z}_{1:n})$. By tracking these positions, we can distinguish the moving target from other stationary scatterers by identifying a scatterer's trajectory that exhibits significant movement. Mathematically, we need to compute the marginal posterior pdf of the augmented state of each PS, which is given by $f(\mathbf{y}_{n,k}|\mathbf{z}_{1:n})=f(\mathbf{x}_{n,k},r_{n,k}|\mathbf{z}_{1:n})$. The posterior existence probability can then be obtained through the following marginalization operation:
\begin{equation}
p(r_{n,k}|\mathbf{z}_{1:n})=\int f(\mathbf{x}_{n,k},r_{n,k}|\mathbf{z}_{1:n})\mathrm{d}\mathbf{x}_{n,k}.
\end{equation}
If the above probability exceeds a threshold $p_{\mathrm{exi}}$, the PS is considered to exist. For an existing scatterer, we then perform the minimum mean square error (MMSE) estimation to determine its position, which is the posterior expectation of the scatterer position \cite{kay1993fundamentals}:
\begin{equation}\label{scatterer estimate}
\hat{\mathbf{x}}_{n,k}=\int \mathbf{x}_{n,k} f(\mathbf{x}_{n,k}|r_{n,k}=1,\mathbf{z}_{1:n})\mathrm{d}\mathbf{x}_{n,k},
\end{equation}
where 
\begin{equation}\label{conditional scatterer}
f(\mathbf{x}_{n,k}|r_{n,k}=1,\mathbf{z}_{1:n})=\frac{f(\mathbf{x}_{n,k},r_{n,k}=1|\mathbf{z}_{1:n})}{p(r_{n,k}=1|\mathbf{z}_{1:n})}.
\end{equation}
Since the transmitter position is also unknown, it needs to be estimated over time as a by-product.
If the marginal posterior pdf of the transmitter position is $f(\mathbf{x}_{n,0}|\mathbf{z}_{1:n})$, 
the position estimate is given by
\begin{equation}
\hat{\mathbf{x}}_{n,0}=\int \mathbf{x}_{n,0} f(\mathbf{x}_{n,0}|\mathbf{z}_{1:n})\mathrm{d}\mathbf{x}_{n,0}.
\end{equation}

Next, we will present the detailed derivation of the joint posterior pdf $f(\mathbf{x}_{1:n,0},\mathbf{y}_{1:n},\mathbf{a}_{1:n}|\mathbf{z}_{1:n})$. To begin this process, we have
\begin{equation}\label{joint post}
\begin{aligned}
&f(\mathbf{x}_{1:n,0},\mathbf{y}_{1:n},\mathbf{a}_{1:n}|\mathbf{z}_{1:n})\\
&=f(\mathbf{x}_{1:n,0},\mathbf{y}_{1:n},\mathbf{a}_{1:n}|\mathbf{z}_{1:n},\mathbf{m}_{1:n})\\
&\propto f(\mathbf{x}_{1:n,0},\mathbf{y}_{1:n},\mathbf{a}_{1:n},\mathbf{z}_{1:n},\mathbf{m}_{1:n})\\
&=f(\mathbf{x}_{1:n,0},\mathbf{y}_{1:n},\mathbf{a}_{1:n},\mathbf{m}_{1:n})f(\mathbf{z}_{1:n}|\mathbf{x}_{1:n,0},\mathbf{y}_{1:n},\mathbf{a}_{1:n},\mathbf{m}_{1:n}),
\end{aligned}
\end{equation}
where in the first step we have used the fact that $M_n$ is known once $\mathbf{z}_n$ is observed. In the following, we will provide the expressions for the two terms in the last step of (\ref{joint post}).

\subsubsection{Derivation of $f(\mathbf{x}_{1:n,0},\mathbf{y}_{1:n},\mathbf{a}_{1:n},\mathbf{m}_{1:n})$}
Assuming that the transmitter position and the augmented states of all legacy PSs evolve independently, we have
\begin{equation}\label{prior all}
\begin{aligned}
&f(\mathbf{x}_{1:n,0},\mathbf{y}_{1:n},\mathbf{a}_{1:n},\mathbf{m}_{1:n})\\
&=\prod_{n'=1}^n 
f(\mathbf{x}_{n',0}|\mathbf{x}_{n'-1,0})\left(\prod_{k=1}^{K_{n'-1}}f(\underline{\mathbf{y}}_{n',k}|\mathbf{y}_{n'-1,k})\right)\\
&~~~~~~~~~\times f(\mathbf{a}_{n'},\overline{\mathbf{y}}_{n'},M_{n'}|\mathbf{x}_{n',0},\underline{\mathbf{y}}_{n'}),
\end{aligned}
\end{equation}
where $f(\underline{\mathbf{y}}_{n,k}|\mathbf{y}_{n-1,k})$ is given in (\ref{transition0}) and (\ref{transition1}). For $n=1$, $f(\mathbf{x}_{1,0}|\mathbf{x}_{0,0})=f(\mathbf{x}_{1,0})$ is the prior pdf of the transmitter's initial position, $f(\underline{\mathbf{y}}_{1,k}|\mathbf{y}_{0,k})$ is set to $1$ because no legacy PS exists at the beginning. The expression of $f(\mathbf{a}_{n},\overline{\mathbf{y}}_{n},M_{n}|\mathbf{x}_{n,0},\underline{\mathbf{y}}_{n})$ is given by
\begin{equation}\label{a y m}
\begin{aligned}
&f(\mathbf{a}_n,\overline{\mathbf{y}}_n,M_n|\mathbf{x}_{n,0},\underline{\mathbf{y}}_n)\\
&=p(\mathbf{a}_n,\overline{\mathbf{r}}_n,M_n|\mathbf{x}_{n,0},\underline{\mathbf{y}}_n)f(\overline{\mathbf{x}}_n|\mathbf{x}_{n,0},\mathbf{a}_n,\overline{\mathbf{r}}_n,M_n,\underline{\mathbf{y}}_n)\\
&=p(\mathbf{a}_n,\overline{\mathbf{r}}_n,M_n|\mathbf{x}_{n,0},\underline{\mathbf{y}}_n)f(\overline{\mathbf{x}}_n|\mathbf{x}_{n,0},\overline{\mathbf{r}}_n,M_n),
\end{aligned}
\end{equation}
where in the second step, we assume that the positions of new PSs are independent of the augmented states of legacy PSs and the association vector. 

 \textbf{Calculation of \bm{$p(\mathbf{a}_n,\overline{\mathbf{r}}_n,M_n|\mathbf{x}_{n,0},\underline{\mathbf{y}}_n)$}:} Following the approach in \cite{meyer2018message,leitinger2019belief}, we assume that the number of measurements generated by new PSs and the number of false alarms are Poisson distributed with mean parameters $\mu_{\mathrm{n},n}$ and $\mu_{\mathrm{FA}}$, respectively. Then $p(\mathbf{a}_n,\overline{\mathbf{r}}_n,M_n|\mathbf{x}_{n,0},\underline{\mathbf{y}}_n)$
can be written as
\begin{equation}\label{a r m pmf}
\begin{aligned}
&p(\mathbf{a}_n,\overline{\mathbf{r}}_n,M_n|\mathbf{x}_{n,0},\underline{\mathbf{y}}_n)\\
&=\frac{1}{M_n!}e^{-(\mu_{\mathrm{FA}}+\mu_{\mathrm{n},n})}(\mu_{\mathrm{n},n})^{\left|\mathcal{N}_{\overline{\mathbf{r}}_n}\right|}
(\mu_{\mathrm{FA}})^{M_n-\left|\mathcal{D}_{\mathbf{a}_n,\underline{\mathbf{r}}_n}\right|-\left|\mathcal{N}_{\overline{\mathbf{r}}_n}\right|}\\
&~~~~\times\Psi(\mathbf{a}_n)\left(\prod_{m\in\mathcal{N}_{\overline{\mathbf{r}}_n}}\Gamma_{\mathbf{a}_n}(\overline{r}_{n,m})\right)
\left(\prod_{k\in\mathcal{D}_{\mathbf{a}_n,\underline{\mathbf{r}}_n}}p_\mathrm{d}(\mathbf{x}_{n,0},\underline{\mathbf{x}}_{n,k})\right)\\
&~~~~\times\prod_{k'\in\bar{\mathcal{D}}_{\mathbf{a}_n,\underline{\mathbf{r}}_n}}\Big(1(a_{n,k'})-\underline{r}_{n,k'}p_\mathrm{d}(\mathbf{x}_{n,0},\underline{\mathbf{x}}_{n,k})\Big),
\end{aligned}
\end{equation}
where $\mathcal{N}_{\overline{\mathbf{r}}_n}\triangleq\left\{m\in\{1,\cdots,M_n\}:\overline{r}_{n,m}=1\right\}$, $\mathcal{D}_{\mathbf{a}_n,\underline{\mathbf{r}}_n}\triangleq\left\{k\in\{1,\cdots,K_{n-1}\}:a_{n,k}\neq 0,~\underline{r}_{n,k}=1\right\}$, $\bar{\mathcal{D}}_{\mathbf{a}_n,\underline{\mathbf{r}}_n}\triangleq\left\{ k\in\{1,\cdots,K_{n-1}\}: k\notin \mathcal{D}_{\mathbf{a}_n,\underline{\mathbf{r}}_n}  \right\}$, and $p_\mathrm{d}(\mathbf{x}_{n,0},\underline{\mathbf{x}}_{n,k})$ is the detection probability of an existing legacy PS, i.e., the probability of an existing legacy PS generating a measurement. $\Psi(\mathbf{a}_n)$, $\Gamma_{\mathbf{a}_n}(\overline{r}_{n,m})$, and $1(a_{n,k})$ are all indicator functions that take values of either $0$ or $1$. Specifically, $\Psi(\mathbf{a}_n)=0$ if there exists $k,k'\in\{1,\cdots,K_{n-1}\}$ with $k\neq k'$ such that $a_{n,k}=a_{n,k'}\neq 0$, and $\Psi(\mathbf{a}_n)=1$ otherwise; $\Gamma_{\mathbf{a}_n}(\overline{r}_{n,m})=0$ if $\overline{r}_{n,m}=1$ and there exists $k\in\{1,\cdots,K_{n-1}\}$ such that $a_{n,k}=m$, and $\Gamma_{\mathbf{a}_n}(\overline{r}_{n,m})=1$ otherwise; $1(a_{n,k})=0$ if $a_{n,k}\neq0$, and $1(a_{n,k})=1$ otherwise. The function $\Psi(\mathbf{a}_n)$ ensures that a measurement cannot be generated simultaneously by more than one legacy PS, and the function $\Gamma_{\mathbf{a}_n}(\overline{r}_{n,m})$ ensures that a measurement cannot be generated simultaneously by a legacy PS and a new PS. We also note that the assumption that each scatterer can generate at most one measurement is implicitly enforced by the structure of the association vector $\mathbf{a}_n$, in which one legacy PS corresponds to only one entry representing the measurement index.
The derivation of (\ref{a r m pmf}) follows similar techniques to those in \cite{meyer2019message} and is omitted in this paper. Then, we further write (\ref{a r m pmf}) as
\begin{equation}\label{a r m pmf 2}
\begin{aligned}
&p(\mathbf{a}_n,\overline{\mathbf{r}}_n,M_n|\mathbf{x}_{n,0},\underline{\mathbf{y}}_n)\\
&=C(M_n)\Psi(\mathbf{a}_n)\left(\prod_{k=1}^{K_{n-1}}g_1(\mathbf{x}_{n,0},\underline{\mathbf{x}}_{n,k},\underline{r}_{n,k},a_{n,k};M_n)\right)\\
&~~~~\times\prod_{m\in\mathcal{N}_{\overline{\mathbf{r}}_n}}\frac{\mu_{\mathrm{n},n}}{\mu_{\mathrm{FA}}}\Gamma_{\mathbf{a}_n}(\overline{r}_{n,m}),
\end{aligned}
\end{equation}
where $C(M_n)$ is a factor that only depends on $M_n$, and $g_1(\mathbf{x}_{n,0},\underline{\mathbf{x}}_{n,k},\underline{r}_{n,k},a_{n,k};M_n)$ is expressed as
\begin{equation}\label{g11}
\begin{aligned}
&g_1(\mathbf{x}_{n,0},\underline{\mathbf{x}}_{n,k},1,a_{n,k};M_n)&\\
&\triangleq
\begin{cases}
    \dfrac{p_\mathrm{d}(\mathbf{x}_{n,0},\underline{\mathbf{x}}_{n,k})}{\mu_\mathrm{FA}},&a_{n,k}=m\in\{1,\cdots,M_n\},\\
    1-p_\mathrm{d}(\mathbf{x}_{n,0},\underline{\mathbf{x}}_{n,k}),&a_{n,k}=0,\\
\end{cases}\\
\end{aligned}
\end{equation}
and
\begin{equation}\label{g12}
g_1(\mathbf{x}_{n,0},\underline{\mathbf{x}}_{n,k},0,a_{n,k};M_n)\triangleq 1(a_{n,k}).    
\end{equation}

\textbf{Calculation of \bm{$f(\overline{\mathbf{x}}_n|\mathbf{x}_{n,0},\overline{\mathbf{r}}_n,M_n)$}:}
Let $f_\mathrm{n}(\overline{\mathbf{x}}_{n,m}|\mathbf{x}_{n,0})$ denote the pdf for the position of an existing new PS and assume that the positions of all existing new PSs are independent and identically distributed. Then, $f(\overline{\mathbf{x}}_n|\mathbf{x}_{n,0},\overline{\mathbf{r}}_n,M_n)$ can be written as
\begin{equation}\label{new PS}
\begin{aligned}
&f(\overline{\mathbf{x}}_{n}|\mathbf{x}_{n,0},\overline{\mathbf{r}}_n,M_n)\\
&=
\left(\prod_{m\in\mathcal{N}_{\overline{\mathbf{r}}_n}}f_\mathrm{n}(\overline{\mathbf{x}}_{n,m}|\mathbf{x}_{n,0})\right)
\left(\prod_{m'\in\bar{\mathcal{N}}_{\overline{\mathbf{r}}_n}}f_\mathrm{D}(\overline{\mathbf{x}}_{n,m'})\right),
\end{aligned}
\end{equation}
where $\bar{\mathcal{N}}_{\overline{\mathbf{r}}_n}\triangleq\left\{m\in\{1,\cdots,M_n\}:m\notin\mathcal{N}_{\overline{\mathbf{r}}_n}\right\}$, and we recall that $f_\mathrm{D}(\overline{\mathbf{x}}_{n,m})$ represents a dummy pdf of $\overline{\mathbf{x}}_{n,m}$.

 Substituting (\ref{a r m pmf 2}) and (\ref{new PS}) back into (\ref{a y m}), we have
\begin{equation}\label{a y m2}
\begin{aligned}
&f(\mathbf{a}_n,\overline{\mathbf{y}}_n,M_n|\mathbf{x}_{n,0},\underline{\mathbf{y}}_n)\\
&=C(M_n)\Psi(\mathbf{a}_n)\left(\prod_{k=1}^{K_{n-1}}g_1(\mathbf{x}_{n,0},\underline{\mathbf{x}}_{n,k},\underline{r}_{n,k},a_{n,k};M_n)\right)\\
&~~~~\times\prod_{m=1}^{M_n}h_1(\mathbf{x}_{n,0},\overline{\mathbf{x}}_{n,m},\overline{r}_{n,m},\mathbf{a}_n),
\end{aligned}
\end{equation}
where $g_1(\mathbf{x}_{n,0},\underline{\mathbf{x}}_{n,k},\underline{r}_{n,k},a_{n,k};M_n)$ is given in $(\ref{g11})$ and $(\ref{g12})$, and $h_1(\mathbf{x}_{n,0},\overline{\mathbf{x}}_{n,m},\overline{r}_{n,m},\mathbf{a}_n)$ is defined as
\begin{equation}\label{h1}
\begin{aligned}
&h_1(\mathbf{x}_{n,0},\overline{\mathbf{x}}_{n,m},1,\mathbf{a}_n)\\&
\triangleq
\begin{cases}
    0,&\exists k\in\{1,\cdots,K_{n-1}\}\\&\text{such~that}~a_{n,k}=m,\\
    \dfrac{\mu_{\mathrm{n},n}}{\mu_{\mathrm{FA}}}f_\mathrm{n}(\overline{\mathbf{x}}_{n,m}|\mathbf{x}_{n,0}),&\text{otherwise},\\
\end{cases}\\
\end{aligned}
\end{equation}
and
\begin{equation}
h_1(\mathbf{x}_{n,0},\overline{\mathbf{x}}_{n,m},0,\mathbf{a}_n)\triangleq f_\mathrm{D}(\overline{\mathbf{x}}_{n,m}).
\end{equation}
For $n=1$, $f(\mathbf{a}_1,\overline{\mathbf{y}}_1,M_1|\mathbf{x}_{1,0},\underline{\mathbf{y}}_1)=f(\overline{\mathbf{y}}_1,M_1|\mathbf{x}_{1,0})$ can be obtained by setting $g_1(\mathbf{x}_{n,0},\underline{\mathbf{x}}_{n,k},\underline{r}_{n,k},a_{n,k};M_n)$ and $\Psi(\mathbf{a}_n)$ in (\ref{a y m2}) to $1$, and modify $h_1(\mathbf{x}_{n,0},\overline{\mathbf{x}}_{n,m},1,\mathbf{a}_n)$ 
 to be $\frac{\mu_{\mathrm{n},n}}{\mu_{\mathrm{FA}}}f_\mathrm{n}(\overline{\mathbf{x}}_{n,m}|\mathbf{x}_{n,0})$, which is independent of $\mathbf{a}_n$. 

\subsubsection{Derivation of $f(\mathbf{z}_{1:n}|\mathbf{x}_{1:n,0},\mathbf{y}_{1:n},\mathbf{a}_{1:n},\mathbf{m}_{1:n})$}
Assuming the conditional independence of measurements across different time instants, we have
\begin{equation}\label{likelihood all}
\begin{aligned}
&f(\mathbf{z}_{1:n}|\mathbf{x}_{1:n,0},\mathbf{y}_{1:n},\mathbf{a}_{1:n},\mathbf{m}_{1:n})\\
&~~~~~~~~~~~~~~=\prod_{n'=1}^n f(\mathbf{z}_{n'}|\mathbf{x}_{n',0},\underline{\mathbf{y}}_{n'},\overline{\mathbf{y}}_{n'},\mathbf{a}_{n'},M_{n'}).
\end{aligned}
\end{equation}
Depending on the origins of the measurements of scattered paths, we can divide them into three categories: those generated by legacy PSs, those generated by new PSs, and false alarms. Therefore, 
$f(\mathbf{z}_{n}|\mathbf{x}_{n,0},\underline{\mathbf{y}}_{n},\overline{\mathbf{y}}_{n},\mathbf{a}_{n},M_{n})$ can be expressed as
\begin{equation}\label{likelihood}
\begin{aligned}
&f(\mathbf{z}_n|\mathbf{x}_{n,0},\underline{\mathbf{y}}_n,\overline{\mathbf{y}}_n,\mathbf{a}_n,M_n)\\
&=\left(\prod_{m=1}^{M_n}f_{\mathrm{FA}}(\mathbf{z}_{n,m})\right)\left(\prod_{k\in\mathcal{D}_{\mathbf{a}_n,\underline{\mathbf{r}}_n}}
\frac{f(\mathbf{z}_{n,a_{n,k}}|\mathbf{x}_{n,0},\underline{\mathbf{x}}_{n,k})}{f_{\mathrm{FA}}(\mathbf{z}_{n,a_{n,k}})}\right)\\
&~~~~\times \left(\prod_{m'\in\mathcal{N}_{\overline{\mathbf{r}}_{n}}}\frac{f(\mathbf{z}_{n,m'}|\mathbf{x}_{n,0},\overline{\mathbf{x}}_{n,m'})}{f_\mathrm{FA}(\mathbf{z}_{n,m'})}\right)f_0(z_{n,0}|\mathbf{x}_{n,0}),
\end{aligned}
\end{equation}
where $f_\mathrm{FA}(\mathbf{z}_{n,m})$ is the pdf of a false alarm measurement, $f(\mathbf{z}_{n,m}|\mathbf{x}_{n,0},\mathbf{x}_{n,k})$ is the conditional pdf of a measurement generated by a true scatterer, and $f_0(z_{n,0}|\mathbf{x}_{n,0})$ is the conditional pdf of the AOA measurement of the direct path. We can further write (\ref{likelihood}) as
\begin{equation}\label{likelihood 2}
\begin{aligned}
&f(\mathbf{z}_n|\mathbf{x}_{n,0},\underline{\mathbf{y}}_n,\overline{\mathbf{y}}_n,\mathbf{a}_n,M_n)\\
&=C(\mathbf{z}_n)f_0(z_{n,0}|\mathbf{x}_{n,0})\left(\prod_{k=1}^{K_{n-1}}g_2(\mathbf{x}_{n,0},\underline{\mathbf{x}}_{n,k},\underline{r}_{n,k},a_{n,k};\mathbf{z}_n)\right)\\
&~~~~\times\prod_{m=1}^{M_n}h_2(\mathbf{x}_{n,0},\overline{\mathbf{x}}_{n,m},\overline{r}_{n,m};\mathbf{z}_{n,m}),
\end{aligned}
\end{equation}
where $C(\mathbf{z}_n)$ is a factor that only depends on $\mathbf{z}_n$ (and also on $M_n$, which can be obtained from $\mathbf{z}_n$), $g_2(\mathbf{x}_{n,0},\underline{\mathbf{x}}_{n,k},\underline{r}_{n,k},a_{n,k};\mathbf{z}_n)$ is expressed as
\begin{equation}\label{g21}
\begin{aligned}
&g_2(\mathbf{x}_{n,0},\underline{\mathbf{x}}_{n,k},1,a_{n,k};\mathbf{z}_n)\\
&\triangleq
\begin{cases}
    \dfrac{f(\mathbf{z}_{n,m}|\mathbf{x}_{n,0},\underline{\mathbf{x}}_{n,k})}{f_\mathrm{FA}(\mathbf{z}_{n,m})},&a_{n,k}=m\in\{1,\cdots,M_n\},\\
    1,&a_{n,k}=0,\\
\end{cases}
\end{aligned}
\end{equation}
and
\begin{equation}\label{g22}
g_2(\mathbf{x}_{n,0},\underline{\mathbf{x}}_{n,k},0,a_{n,k};\mathbf{z}_n)\triangleq1,
\end{equation}
and $h_2(\mathbf{x}_{n,0},\overline{\mathbf{x}}_{n,m},\overline{r}_{n,m};\mathbf{z}_{n,m})$ is expressed as
\begin{equation}\label{h21}
\begin{aligned}
&h_2(\mathbf{x}_{n,0},\overline{\mathbf{x}}_{n,m},\overline{r}_{n,m};\mathbf{z}_{n,m})\\
&~~~~~~~~~~~~~~~~~\triangleq
\begin{cases}
    \dfrac{f(\mathbf{z}_{n,m}|\mathbf{x}_{n,0},\overline{\mathbf{x}}_{n,m})}{f_\mathrm{FA}(\mathbf{z}_{n,m})},&\overline{r}_{n,m}=1,\\
    1,&\overline{r}_{n,m}=0.\\
\end{cases}
\end{aligned}
\end{equation}
For $n=1$, we have $f(\mathbf{z}_{1}|\mathbf{x}_{1,0},\underline{\mathbf{y}}_{1},\overline{\mathbf{y}}_{1},\mathbf{a}_{1},M_{1})=f(\mathbf{z}_{1}|\mathbf{x}_{1,0},\overline{\mathbf{y}}_{1},M_{1})$, which can be obtained by setting the product involving $g_2(\mathbf{x}_{n,0},\underline{\mathbf{x}}_{n,k},\underline{r}_{n,k},a_{n,k};\mathbf{z}_n)$ in (\ref{likelihood 2}) to $1$.

Substituting the detailed expressions of (\ref{prior all}) and (\ref{likelihood all}) back into (\ref{joint post}), we have
\begin{equation}\label{joint 1}
\begin{aligned}
&f(\mathbf{x}_{1:n,0},\mathbf{y}_{1:n},\mathbf{a}_{1:n}|\mathbf{z}_{1:n})\\
&\propto\prod_{n'=1}^nf(\mathbf{x}_{n',0}|\mathbf{x}_{n'-1,0})f_0(z_{n',0}|\mathbf{x}_{n',0})\Psi(\mathbf{a}_{n'})\\
&~~~~\times\left(\prod_{k=1}^{K_{n'-1}}f(\underline{\mathbf{y}}_{n',k}|\mathbf{y}_{n'-1,k})g(\mathbf{x}_{n',0},\mathbf{\underline{x}}_{n',k},\underline{r}_{n',k},a_{n',k};\mathbf{z}_{n'})\right)\\
&~~~~\times\left(\prod_{m=1}^{M_{n'}}h_1(\mathbf{x}_{n',0},\overline{\mathbf{x}}_{n',m},\overline{r}_{n',m},\mathbf{a}_{n'})\right.\\
&~~~~~~~~~~~~~~~\times h_2(\mathbf{x}_{n',0},\overline{\mathbf{x}}_{n',m},\overline{r}_{n',m};\mathbf{z}_{n',m})\Bigg),
\end{aligned}
\end{equation}
where $g(\mathbf{x}_{n,0},\underline{\mathbf{x}}_{n,k},\underline{r}_{n,k},a_{n,k};\mathbf{z}_{n})$ is the product of $g_1(\mathbf{x}_{n,0},\underline{\mathbf{x}}_{n,k},\underline{r}_{n,k},a_{n,k};M_{n})$ and $g_2(\mathbf{x}_{n,0},\underline{\mathbf{x}}_{n,k},\underline{r}_{n,k},a_{n,k};\mathbf{z}_{n})$, which is expressed as
\begin{equation}\label{g1}
\begin{aligned}
&g(\mathbf{x}_{n,0},\underline{\mathbf{x}}_{n,k},1,a_{n,k};\mathbf{z}_{n})\\
 &=\begin{cases}
     \dfrac{p_\mathrm{d}(\mathbf{x}_{n,0},\underline{\mathbf{x}}_{n,k}) f(\mathbf{z}_{n,m}|\mathbf{x}_{n,0},\underline{\mathbf{x}}_{n,k})}{\mu_{\mathrm{FA}}f_\mathrm{FA}(\mathbf{z}_{n,m})},
     &a_{n,k}=m\\
     &\in\{1,\cdots,M_{n}\},\\
     1-p_\mathrm{d}(\mathbf{x}_{n,0},\underline{\mathbf{x}}_{n,k}),&a_{n,k}=0,\\
 \end{cases}
\end{aligned}
\end{equation}
and
\begin{equation}\label{g0}
g(\mathbf{x}_{n,0},\underline{\mathbf{x}}_{n,k},0,a_{n,k};\mathbf{z}_{n})=1(a_{n,k}).
\end{equation}
Since $\Psi(\mathbf{a}_{n})$ is a function of $\mathbf{a}_{n}$, with $\mathbf{a}_{n}$ taking values from the $K_{n-1}$-fold Cartesian product of $\{0,\cdots,M_n\}$, the computational complexity grows exponentially with $K_{n-1}$. To address this challenge, we adopt the approach in  \cite{williams2010data,williams2014approximate,meyer2018message,leitinger2019belief} and introduce a new function of both $\mathbf{a}_{n}$ and $\mathbf{b}_n$ that can be further ``opened'':
\begin{equation}
\Psi(\mathbf{a}_n,\mathbf{b}_n)=\prod_{k=1}^{K_{n-1}}\prod_{m=1}^{M_n}\psi(a_{n,k},b_{n,m}),
\end{equation}
where
\begin{equation}
\begin{aligned}
\psi(a_{n,k},b_{n,m})=
\begin{cases}
    0,&a_{n,k}=m,b_{n,m}\neq k\\&~~~~~~~~\text{or}~b_{n,m}=k,a_{n,k}\neq m,\\
    1,&\text{otherwise}.\\
\end{cases}
\end{aligned}
\end{equation}
With this formulation, we can observe that the complexity for each $\psi(a_{n,k},b_{n,m})$ grows linearly with $K_{n-1}$ and $M_n$, which will be utilized in the proposed target tracking algorithm. Denote $\mathbf{b}_{1:n}\triangleq[\mathbf{b}_1^\mathrm{T},\cdots,\mathbf{b}_n^\mathrm{T}]^\mathrm{T}$. We can now derive the new posterior pdf $f(\mathbf{x}_{1:n,0},\mathbf{y}_{1:n},\mathbf{a}_{1:n},\mathbf{b}_{1:n}|\mathbf{z}_{1:n})$ by replacing $\Psi(\mathbf{a}_{n'})$ in (\ref{joint 1}) with $\Psi(\mathbf{a}_{n'},\mathbf{b}_{n'})$. In addition, the product of $h_1(\mathbf{x}_{n,0},\overline{\mathbf{x}}_{n,m},\overline{r}_{n,m},\mathbf{a}_{n})$ and $h_2(\mathbf{x}_{n,0},\overline{\mathbf{x}}_{n,m},\overline{r}_{n,m};\mathbf{z}_{n,m})$ can be replaced by a new factor $h(\mathbf{x}_{n,0},\overline{\mathbf{x}}_{n,m},\overline{r}_{n,m},b_{n,m};\mathbf{z}_{n,m})$ expressed as
\begin{equation}\label{h1}
\small
\begin{aligned}
&h(\mathbf{x}_{n,0},\overline{\mathbf{x}}_{n,m},1,b_{n,m};\mathbf{z}_{n,m})\\
&\triangleq
\begin{cases}
    0,&b_{n,m}=k\\&\in\{1,\cdots,K_{n-1}\},\\
    \dfrac{\mu_{\mathrm{n},n}f_\mathrm{n}(\overline{\mathbf{x}}_{n,m}|\mathbf{x}_{n,0})f(\mathbf{z}_{n,m}|\mathbf{x}_{n,0},\overline{\mathbf{x}}_{n,m})}{\mu_{\mathrm{FA}}f_{\mathrm{FA}}(\mathbf{z}_{n,m})},&b_{n,m}=0,\\
\end{cases}
\end{aligned}
\end{equation}
and
\begin{equation}\label{h0}
h(\mathbf{x}_{n,0},\overline{\mathbf{x}}_{n,m},0,b_{n,m};\mathbf{z}_{n,m})\triangleq f_\mathrm{D}(\overline{\mathbf{x}}_{n,m}).
\end{equation}
Therefore, the new posterior pdf is given by
\begin{equation}\label{joint 2}
\begin{aligned}
&f(\mathbf{x}_{1:n,0},\mathbf{y}_{1:n},\mathbf{a}_{1:n},\mathbf{b}_{1:n}|\mathbf{z}_{1:n})\\
&\propto\prod_{n'=1}^nf(\mathbf{x}_{n',0}|\mathbf{x}_{n'-1,0})f_0(z_{n',0}|\mathbf{x}_{n',0})\\
&~~~~\times\left(\prod_{k=1}^{K_{n'-1}}f(\underline{\mathbf{y}}_{n',k}|\mathbf{y}_{n'-1,k})g(\mathbf{x}_{n',0},\mathbf{\underline{x}}_{n',k},\underline{r}_{n',k},a_{n',k};\mathbf{z}_{n'})\right.\\
&~~~~~~~~~~~~~~~~\times\prod_{m'=1}^{M_{n'}}\psi(a_{n',k},b_{n',m'})\Bigg)\\
&~~~~\times\prod_{m=1}^{M_{n'}}h(\mathbf{x}_{n',0},\overline{\mathbf{x}}_{n',m},\overline{r}_{n',m},b_{n',m};\mathbf{z}_{n',m}).
\end{aligned}
\end{equation}

However, it is a nontrivial task to compute the marginal posterior distribution of any variable from the joint distribution in (\ref{joint 2}). In the next section, we will present a low-complexity belief propagation-based method to approximate these marginal distributions.
This algorithm adapts the SLAM framework in \cite{leitinger2019belief}, where a primary modification is that after predicting the transmitter position at each time instant, we further refine it using the AOA measurement of the direct path. In other words, this step combines the predicted pdf of $\mathbf{x}_{n,0}$ and the measurement model $f_0(z_{n,0}|\mathbf{x}_{n,0})$ for the direct path to form a composite prediction step. The combination is represented as additional messages on the factor graph. 
Another difference from the implementation perspective is that our target tracking algorithm uses the relative distance and AOA as measurements of scatterers, whereas \cite{leitinger2019belief} uses only the absolute distance. The AOA introduces ambiguity since it is unclear from which side the signal arrives at the receiver's antenna array. Consequently, we need to account for this ambiguity during the initialization of the transmitter and scatterer positions in our simulations.

\section{Belief Propagation for Passive Target Tracking}\label{bp algorithm}

\begin{figure}
\centering
\includegraphics[width=1\columnwidth]{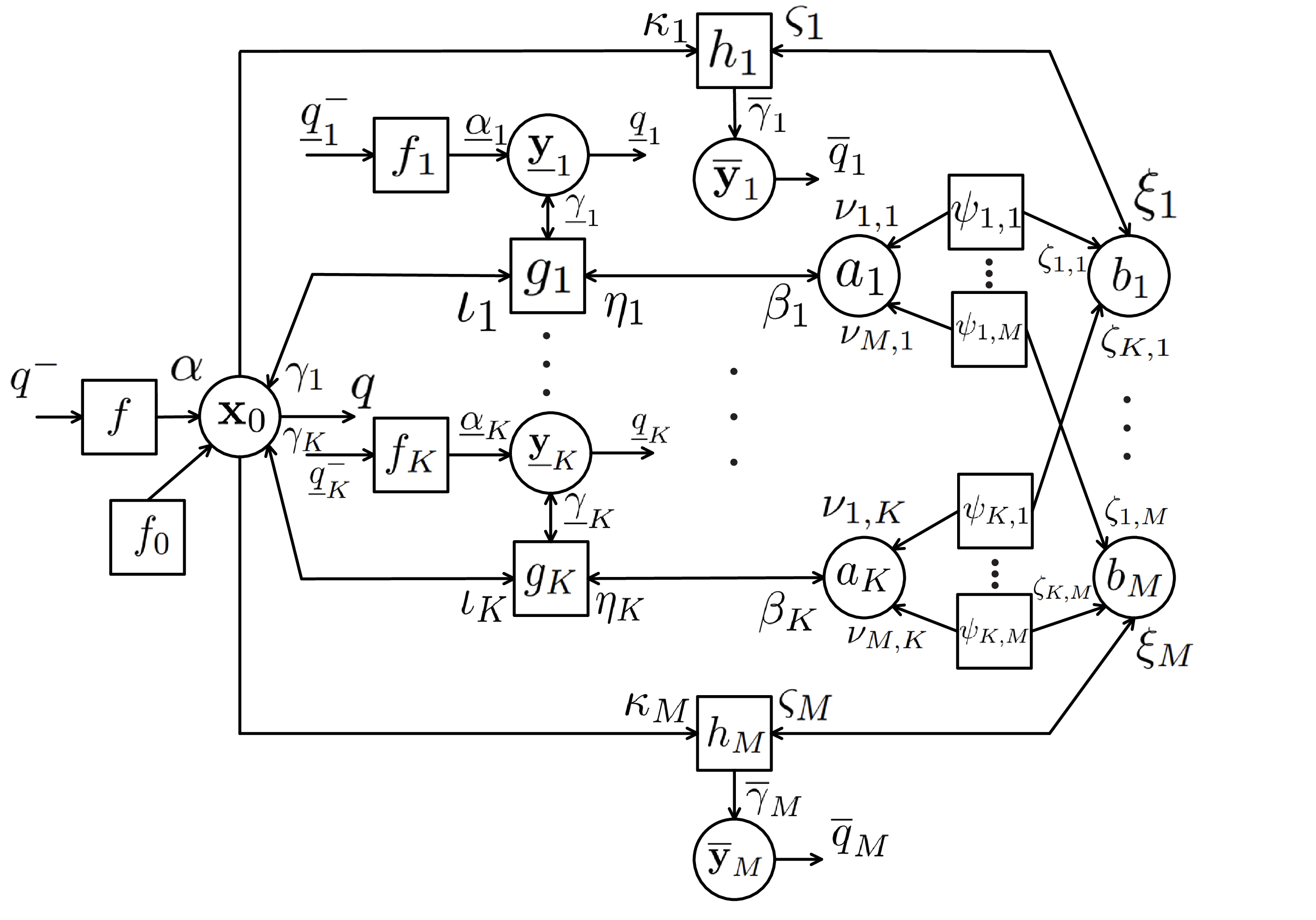}
\caption{Factor graph of the posterior pdf in (\ref{joint 2}) for a single time instant $n$, in which factor nodes are represented by squares, variable nodes by circles, and messages passed on edges by directed arrows. For simplicity, we omit the time index $n$ for the variables in circles and use the following shorthand notations: $M\triangleq M_n$, $K\triangleq K_{n-1}$, $f\triangleq f(\mathbf{x}_{n,0}|\mathbf{x}_{n-1,0})$, $f_k\triangleq f(\underline{\mathbf{y}}_{n,k}|\mathbf{y}_{n-1,k})$, $f_0\triangleq f_0(z_{n,0}|\mathbf{x}_{n,0})$, $g_k\triangleq g(\mathbf{x}_{n,0},\underline{\mathbf{x}}_{n,k},\underline{r}_{n,k},a_{n,k};\mathbf{z}_n)$, $h_m\triangleq h(\mathbf{x}_{n,0},\overline{\mathbf{x}}_{n,m},\overline{r}_{n,m},b_{n,m};\mathbf{z}_{n,m})$, $\psi_{k,m}\triangleq \psi(a_{n,k},b_{n,m})$, $q^-\triangleq q(\mathbf{x}_{n-1,0})$, $q\triangleq q(\mathbf{x}_{n,0})$, $\underline{q}^-_k\triangleq \underline{q}(\mathbf{y}_{n-1,k})$, $\underline{q}_k\triangleq \underline{q}(\underline{\mathbf{y}}_{n,k})$, $\overline{q}_m\triangleq \overline{q}(\overline{\mathbf{y}}_{n,m})$, $\alpha\triangleq \alpha(\mathbf{x}_{n,0})$, $\underline{\alpha}_k\triangleq \underline{\alpha}(\underline{\mathbf{y}}_{n,k})$, $\iota_k\triangleq\iota_k(\mathbf{x}_{n,0})$, $\kappa_m\triangleq \kappa_m(\mathbf{x}_{n,0})$, $\beta_k\triangleq\beta(a_{n,k})$, $\xi_m\triangleq\xi(b_{n,m})$, $\nu_{m,k}\triangleq \nu_{m\rightarrow k}^{(p)}(a_{n,k})$, $\zeta_{k,m}\triangleq\zeta_{k\rightarrow m}^{(p)}(b_{n,m})$, $\eta_k\triangleq\eta(a_{n,k})$, $\varsigma_m\triangleq\varsigma(b_{n,m})$, $\gamma_k\triangleq\gamma_k(\mathbf{x}_{n,0})$,  $\underline{\gamma}_k\triangleq\underline{\gamma}(\underline{\mathbf{y}}_{n,k})$, and $\overline{\gamma}_m\triangleq\overline{\gamma}(\overline{\mathbf{y}}_{n,m})$.
}
\label{localization model}
\end{figure}

Belief propagation is an efficient method for computing marginal distributions of variables in a complicated probabilistic model \cite{kschischang2001factor,loeliger2004introduction,yedidia2005constructing}. In this method, the joint distribution is represented by a factor graph consisting of factor nodes, variable nodes, and edges. On the factor graph, messages are passed on edges between variables and factors, and the marginal distribution of a variable is computed as the product of all incoming messages of the variable. In Fig. \ref{localization model}, we depict the factor graph corresponding to the pdf in (\ref{joint 2}) for a single time instant. Since the factor graph contains loops, we need to specify an order for the message computation. As in \cite{meyer2017scalable,leitinger2019belief}, we only compute messages passed forward in time. This rule indicates that later beliefs are not used to refine former estimation results, thereby facilitating real-time applications. In addition, we only perform iterative message passing at each time instant for the association between legacy PSs and measurements. Although incorporating more loops and iterations could potentially improve the overall performance by allowing additional \textit{probabilistic information transfer} \cite{meyer2015distributed}, we choose not to include them in order to maintain a low computational complexity. Based on the message passing rules defined above, we will next derive the expressions for all the messages.

\subsubsection{Prediction for the transmitter position} A prediction step, using the belief from the previous time instant $q(\mathbf{x}_{n-1,0})$ and the transition pdf $f(\mathbf{x}_{n,0}|\mathbf{x}_{n-1,0})$, is performed for the transmitter position. This step corresponds to the message passed from the factor $f(\mathbf{x}_{n,0}|\mathbf{x}_{n-1,0})$ to the variable $\mathbf{x}_{n,0}$, which can be calculated as
\begin{equation}\label{message alpha}
\alpha(\mathbf{x}_{n,0})=\int f(\mathbf{x}_{n,0}|\mathbf{x}_{n-1,0})q(\mathbf{x}_{n-1,0})\mathrm{d}\mathbf{x}_{n-1,0}.
\end{equation}
It can be shown that if $q(\mathbf{x}_{n-1,0})$ is normalized such that it integrates to $1$, then $\alpha(\mathbf{x}_{n,0})$ is also normalized. 

\subsubsection{Prediction for the augmented states of legacy PSs} Similar to above, we perform the prediction step for each legacy PS. The message passed from the factor $f(\underline{\mathbf{y}}_{n,k}|\mathbf{y}_{n-1,k})$ to the variable $\underline{\mathbf{y}}_{n,k}$ is given by
\begin{equation}\label{predict PS}
\begin{aligned}
&\underline{\alpha}(\underline{\mathbf{x}}_{n,k},\underline{r}_{n,k})=\sum_{r_{n-1,k}\in\{0,1\}}\int f(\underline{\mathbf{x}}_{n,k},\underline{r}_{n,k}|\mathbf{x}_{n-1,k},r_{n-1,k})\\
&~~~~~~~~~~~~~~~~~~~~~~~~~~~~~~~~~~~~\times\underline{q}(\mathbf{x}_{n-1,k},r_{n-1,k})\mathrm{d}\mathbf{x}_{n-1,k},
\end{aligned}
\end{equation}
where $\underline{q}(\mathbf{x}_{n-1,k},r_{n-1,k})$ is the belief of the $k$-th PS at time $n-1$. Substituting (\ref{transition0}) and (\ref{transition1}) into (\ref{predict PS}), we have
\begin{equation}\label{predict PS0}
\begin{aligned}
&\underline{\alpha}(\underline{\mathbf{x}}_{n,k},0)=f_\mathrm{D}(\underline{\mathbf{x}}_{n,k})\int\underline{q}(\mathbf{x}_{n-1,k},0)\mathrm{d}\mathbf{x}_{n-1,k}\\
&~~~~~~~~~~~~~~~~+(1-p_\mathrm{s})f_\mathrm{D}(\underline{\mathbf{x}}_{n,k})\int\underline{q}(\mathbf{x}_{n-1,k},1)\mathrm{d}\mathbf{x}_{n-1,k},
\end{aligned}
\end{equation}
and
\begin{equation}\label{predict PS1}
\begin{aligned}
\underline{\alpha}(\underline{\mathbf{x}}_{n,k},1)=p_\mathrm{s}\int f(\underline{\mathbf{x}}_{n,k}|\mathbf{x}_{n-1,k})\underline{q}(\mathbf{x}_{n-1,k},1)\mathrm{d}\mathbf{x}_{n-1,k}.
\end{aligned}
\end{equation}
Denote $\underline{\alpha}_{n,k}\triangleq\int\underline{\alpha}(\underline{\mathbf{x}}_{n,k},0)\mathrm{d}\underline{\mathbf{x}}_{n,k}$ and $\underline{q}_{n-1,k}\triangleq\int \underline{q}(\mathbf{x}_{n-1,k},0)\mathrm{d}\mathbf{x}_{n-1,k}$, which can be interpreted as the predicted non-existence probability at time $n$ and the posterior non-existence probability at time $n-1$, respectively. Based on (\ref{predict PS0}), we can obtain
\begin{equation}\label{predict probability ps0}
\underline{\alpha}_{n,k}=\underline{q}_{n-1,k}+(1-p_\mathrm{s})\int\underline{q}(\mathbf{x}_{n-1,k},1)\mathrm{d}\mathbf{x}_{n-1,k},
\end{equation}
where we have used the assumption that the dummy pdf $f_\mathrm{D}(\underline{\mathbf{x}}_{n,k})$ integrates to $1$. This equation will be utilized in subsequent message computations. Similar to $\alpha(\mathbf{x}_{n,0})$, if $\underline{q}(\mathbf{x}_{n-1,k},r_{n-1,k})$ is normalized such that $\sum_{r_{n-1,k}\in\{0,1\}}\int\underline{q}(\mathbf{x}_{n-1,k},r_{n-1,k})\mathrm{d}\mathbf{x}_{n-1,k}=1$, then $\underline{\alpha}(\underline{\mathbf{x}}_{n,k},\underline{r}_{n,k})$ is also normalized.

\subsubsection{Measurement evaluation for the transmitter position} In this step, we compute the messages passed from the variable $\mathbf{x}_{n,0}$ to the factors $g(\mathbf{x}_{n,0},\underline{\mathbf{x}}_{n,k},\underline{r}_{n,k},a_{n,k};\mathbf{z}_n)$ and $h(\mathbf{x}_{n,0},\overline{\mathbf{x}}_{n,m},\overline{r}_{n,m},b_{n,m};\mathbf{z}_{n,m})$. For all $k\in\{1,\cdots,K_{n-1}\}$ and $m\in\{1,\cdots,M_n\}$, we have
\begin{equation}\label{iota kappa}
\iota_k(\mathbf{x}_{n,0})=\kappa_m(\mathbf{x}_{n,0})\propto\alpha(\mathbf{x}_{n,0})f_0(z_{n,0}|\mathbf{x}_{n,0}),
\end{equation}
where we use ``$\propto$'' instead of ``$=$'' in the second step because the messages need to be normalized such that they integrate to $1$, which facilitates the subsequent derivations. As mentioned earlier, we only consider the loop in the factor graph for the association between legacy PSs and measurements, and therefore, for each $\iota_k(\mathbf{x}_{n,0})$ or $\kappa_m(\mathbf{x}_{n,0})$, we have not taken into account the messages passed from factors corresponding to other PSs back to $\mathbf{x}_{n,0}$.

\subsubsection{Measurement evaluation for the augmented states of legacy PSs} The message passed from the factor $g(\mathbf{x}_{n,0},\underline{\mathbf{x}}_{n,k},\underline{r}_{n,k},a_{n,k};\mathbf{z}_n)$ to the variable $a_{n,k}$ is computed as
\begin{equation}\label{message beta}
\begin{aligned}
\beta(a_{n,k})&=\sum_{\underline{r}_{n,k}\in\{0,1\}}\iint g(\mathbf{x}_{n,0},\underline{\mathbf{x}}_{n,k},\underline{r}_{n,k},a_{n,k};\mathbf{z}_n)\\
&~~~~~~~\times\iota_k(\mathbf{x}_{n,0})\underline{\alpha}(\underline{\mathbf{x}}_{n,k},\underline{r}_{n,k})\mathrm{d}\mathbf{x}_{n,0}\mathrm{d}\underline{\mathbf{x}}_{n,k}\\
&=\iint g(\mathbf{x}_{n,0},\underline{\mathbf{x}}_{n,k},1,a_{n,k};\mathbf{z}_n)\iota_k(\mathbf{x}_{n,0})\\
&~~~~~~~\times\underline{\alpha}(\underline{\mathbf{x}}_{n,k},1)\mathrm{d}\mathbf{x}_{n,0}\mathrm{d}\underline{\mathbf{x}}_{n,k}+1(a_{n,k})\underline{\alpha}_{n,k},
\end{aligned}
\end{equation}
where in the second step we have used the expressions in (\ref{g1})--(\ref{g0}) and the fact that $\iota_k(\mathbf{x}_{n,0})$ integrates to $1$.

\subsubsection{Measurement evaluation for the augmented states of new PSs} The message passed from the factor $h(\mathbf{x}_{n,0},\overline{\mathbf{x}}_{n,m},\overline{r}_{n,m},b_{n,m};\mathbf{z}_{n,m})$ to the variable $b_{n,m}$ is computed as
\begin{equation}\label{message xi}
\begin{aligned}
\xi(b_{n,m})&=\sum_{\overline{r}_{n,m}\in\{0,1\}}\iint h(\mathbf{x}_{n,0},\overline{\mathbf{x}}_{n,m},\overline{r}_{n,m},b_{n,m};\mathbf{z}_{n,m})\\
&~~~~~~~~~~~~~~~~~~~~\times\kappa_m(\mathbf{x}_{n,0})\mathrm{d}\mathbf{x}_{n,0}\mathrm{d}\overline{\mathbf{x}}_{n,m}.
\end{aligned}
\end{equation}
Substituting the expressions for $h(\mathbf{x}_{n,0},\overline{\mathbf{x}}_{n,m},\overline{r}_{n,m},b_{n,m};\mathbf{z}_{n,m})$ from (\ref{h1}) and (\ref{h0}) into the above equation, we have $\xi(b_{n,m})=1$ for $b_{n,m}\in\{1,\cdots,K_{n-1}\}$, and
\begin{equation}\label{message xi 0}
\begin{aligned}
\xi(b_{n,m})=&1+\frac{\mu_{\mathrm{n},n}}{\mu_\mathrm{FA}f_\mathrm{FA}(\mathbf{z}_{n,m})}\iint f_\mathrm{n}(\overline{\mathbf{x}}_{n,m}|\mathbf{x}_{n,0})\\
&\times f(\mathbf{z}_{n,m}|\mathbf{x}_{n,0},\overline{\mathbf{x}}_{n,m})\kappa_m(\mathbf{x}_{n,0})\mathrm{d}\mathbf{x}_{n,0}\mathrm{d}\overline{\mathbf{x}}_{n,m}
\end{aligned}
\end{equation}
for $b_{n,m}=0$, where we have used the fact that both $f_\mathrm{D}(\overline{\mathbf{x}}_{n,m})$ 
and $\kappa_m(\mathbf{x}_{n,0})$ integrate to $1$.

\subsubsection{Association between legacy PSs and measurements} Iterative message passing is performed for the association between legacy PSs and measurements. In the $p$-th iteration, let $\nu_{m\rightarrow k}^{(p)}(a_{n,k})$ denote the message passed from the factor $\psi(a_{n,k},b_{n,m})$ to the variable $a_{n,k}$, and $\zeta_{k\rightarrow m}^{(p)}(b_{n,m})$ denote the message passed from the factor $\psi(a_{n,k},b_{n,m})$ to the variable $b_{n,m}$. Then we have
\begin{equation}\label{iterative1}
\begin{aligned}
&\nu_{m\rightarrow k}^{(p)}(a_{n,k})\\
&=\sum_{b_{n,m}=0}^{K_{n-1}}\psi(a_{n,k},b_{n,m})\xi(b_{n,m})\prod_{k'=1,k'\neq k}^{K_{n-1}}\zeta_{k'\rightarrow m}^{(p-1)}(b_{n,m}),
\end{aligned}
\end{equation}
and
\begin{equation}\label{iterative2}
\begin{aligned}
&\zeta_{k\rightarrow m}^{(p)}(b_{n,m})\\
&=\sum_{a_{n,k}=0}^{M_n}\psi(a_{n,k},b_{n,m})\beta(a_{n,k})\prod_{m'=1,m'\neq m}^{M_n}\nu_{m'\rightarrow k}^{(p)}(a_{n,k}),
\end{aligned}
\end{equation}
where $p\in\{1,\cdots,P\}$. To start the iteration, we initialize $\zeta_{k'\rightarrow m}^{(0)}(b_{n,m})=1$. After the iteration converges, the message passed from $a_{n,k}$ back to $g(\mathbf{x}_{n,0},\underline{\mathbf{x}}_{n,k},\underline{r}_{n,k},a_{n,k};\mathbf{z}_n)$ is given by
\begin{equation}\label{message eta}
    \eta(a_{n,k})=\prod_{m=1}^{M_n}\nu_{m\rightarrow k}^{(P)}(a_{n,k}),
\end{equation}
and the message passed from $b_{n,m}$ back to $h(\mathbf{x}_{n,0},\overline{\mathbf{x}}_{n,m},\overline{r}_{n,m},b_{n,m};\mathbf{z}_{n,m})$ is given by
\begin{equation}\label{message varsigma}
\varsigma(b_{n,m})=\prod_{k=1}^{K_{n-1}}\zeta_{k\rightarrow m}^{(P)}(b_{n,m}).
\end{equation}

\subsubsection{Measurement update for the transmitter position} The message from the factor $g(\mathbf{x}_{n,0},\underline{\mathbf{x}}_{n,k},\underline{r}_{n,k},a_{n,k};\mathbf{z}_n)$ back to the variable $\mathbf{x}_{n,0}$ is computed as
\begin{equation}\label{message gamma_k}
\begin{aligned}
\gamma_k(\mathbf{x}_{n,0})&=\sum_{\underline{r}_{n,k}\in\{0,1\}}\sum_{a_{n,k}=0}^{M_n}\int g(\mathbf{x}_{n,0},\underline{\mathbf{x}}_{n,k},\underline{r}_{n,k},a_{n,k};\mathbf{z}_n)\\
&~~~~~\times\underline{\alpha}(\underline{\mathbf{x}}_{n,k},\underline{r}_{n,k})\eta(a_{n,k})\mathrm{d}\underline{\mathbf{x}}_{n,k}\\
&=\sum_{a_{n,k}=0}^{M_n}\eta(a_{n,k})\int g(\mathbf{x}_{n,0},\underline{\mathbf{x}}_{n,k},1,a_{n,k};\mathbf{z}_n)\\
&~~~~~\times\underline{\alpha}(\underline{\mathbf{x}}_{n,k},1)\mathrm{d}\underline{\mathbf{x}}_{n,k}+\eta(0)\underline{\alpha}_{n,k}.
\end{aligned}
\end{equation}
As in \cite{leitinger2019belief}, we do not consider the messages from $h(\mathbf{x}_{n,0},\overline{\mathbf{x}}_{n,m},\overline{r}_{n,m},b_{n,m};\mathbf{z}_{n,m})$ back to $\mathbf{x}_{n,0}$, as these messages associated with new PSs may not provide sufficient reliable information for updating the belief of the transmitter position at a later stage.

\subsubsection{Measurement update for the augmented states of legacy PSs} Similar to above, the message from the factor $g(\mathbf{x}_{n,0},\underline{\mathbf{x}}_{n,k},\underline{r}_{n,k},a_{n,k};\mathbf{z}_n)$ to the variable $\underline{\mathbf{y}}_{n,k}$ is given by
\begin{equation}\label{message legacy gamma 1}
\begin{aligned}
&\underline{\gamma}(\underline{\mathbf{x}}_{n,k},1)\\
&=\sum_{a_{n,k}=0}^{M_n}\eta(a_{n,k})\int g(\mathbf{x}_{n,0},\underline{\mathbf{x}}_{n,k},1,a_{n,k};\mathbf{z}_n)\iota_k(\mathbf{x}_{n,0})\mathrm{d}\mathbf{x}_{n,0},
\end{aligned}
\end{equation}
and
\begin{equation}\label{message legacy gamma 0}
\underline{\gamma}(\underline{\mathbf{x}}_{n,k},0)=\eta(0).
\end{equation}

\subsubsection{Measurement update for the augmented states of new PSs} The message from the factor $h(\mathbf{x}_{n,0},\overline{\mathbf{x}}_{n,m},\overline{r}_{n,m},b_{n,m};\mathbf{z}_{n,m})$ to the variable $\overline{\mathbf{y}}_{n,m}$ is given by
\begin{equation}\label{message new phi 1}
\begin{aligned}
\overline{\gamma}(\overline{\mathbf{x}}_{n,m},1)=\varsigma(0)\int h(\mathbf{x}_{n,0},\overline{\mathbf{x}}_{n,m},1,0;\mathbf{z}_{n,m})\kappa_m(\mathbf{x}_{n,0})\mathrm{d}\mathbf{x}_{n,0},
\end{aligned}
\end{equation}
and
\begin{equation}\label{message new phi 0}
\begin{aligned}
\overline{\gamma}(\overline{\mathbf{x}}_{n,m},0)=\sum_{b_{n,m}=0}^{K_{n-1}}f_\mathrm{D}(\overline{\mathbf{x}}_{n,m})\varsigma(b_{n,m}).
\end{aligned}
\end{equation}
Denote $\overline{\gamma}_{n,m}\triangleq\int\overline{\gamma}(\overline{\mathbf{x}}_{n,m},0)\mathrm{d}\overline{\mathbf{x}}_{n,m}$. We then have $\overline{\gamma}_{n,m}=\sum_{b_{n,m}=0}^{K_{n-1}}\varsigma(b_{n,m})$.

\subsubsection{Belief for the transmitter position} The belief for the transmitter position, which serves as an approximation to its posterior pdf, is computed as
\begin{equation}\label{illuminator belief}
f(\mathbf{x}_{n,0}|\mathbf{z}_{1:n})\approx q(\mathbf{x}_{n,0})\propto\alpha(\mathbf{x}_{n,0})f_0(z_{n,0}|\mathbf{x}_{n,0})\prod_{k=1}^{K_{n-1}}\gamma_k(\mathbf{x}_{n,0}),
\end{equation}
where we use ``$\propto$'' in the second step because the belief is normalized such that it integrates to $1$. 

\subsubsection{Belief for the augmented states of legacy PSs} The belief for the augmented state of each legacy PS is computed as
\begin{equation}\label{legacy belief}
\begin{aligned}
f(\underline{\mathbf{x}}_{n,k},\underline{r}_{n,k}|\mathbf{z}_{1:n})&\approx \underline{q}(\underline{\mathbf{x}}_{n,k},\underline{r}_{n,k})\\
&\propto\underline{\alpha}(\underline{\mathbf{x}}_{n,k},\underline{r}_{n,k})\underline{\gamma}(\underline{\mathbf{x}}_{n,k},\underline{r}_{n,k}),
\end{aligned}
\end{equation}
and the posterior non-existence probability is given by
\begin{equation}\label{legacy nonexistence}
p(\underline{r}_{n,k}=0|\mathbf{z}_{1:n})\approx\underline{q}_{n,k}\triangleq \int \underline{q}(\underline{\mathbf{x}}_{n,k},0)\mathrm{d}\underline{\mathbf{x}}_{n,k}\propto\eta(0)\underline{\alpha}_{n,k},
\end{equation}
where we use ``$\propto$'' because the belief is normalized, with the normalization constant given by
\begin{equation}\label{normalization constant legacy}
\underline{C}_{n,k}=\frac{1}{\int \underline{\alpha}(\underline{\mathbf{x}}_{n,k},1)\underline{\gamma}(\underline{\mathbf{x}}_{n,k},1)\mathrm{d}\underline{\mathbf{x}}_{n,k}+\eta(0)\underline{\alpha}_{n,k}}.
\end{equation}

\subsubsection{Belief for the augmented states of new PSs} The belief for the augmented state of each new PS is computed as
\begin{equation}\label{new belief}
\begin{aligned}
f(\overline{\mathbf{x}}_{n,m},\overline{r}_{n,m}|\mathbf{z}_{1:n})&\approx \overline{q}(\overline{\mathbf{x}}_{n,m},\overline{r}_{n,m})\propto\overline{\gamma}(\overline{\mathbf{x}}_{n,m},\overline{r}_{n,m}),
\end{aligned}
\end{equation}
and the posterior non-existence probability is given by
\begin{equation}
p(\overline{r}_{n,m}=0|\mathbf{z}_{1:n})\approx\overline{q}_{n,m}\triangleq \int \overline{q}(\overline{\mathbf{x}}_{n,m},0)\mathrm{d}\overline{\mathbf{x}}_{n,m}\propto\overline{\gamma}_{n,m},
\end{equation}
where we normalize the belief with the normalization constant given by
\begin{equation}\label{normalization constant new}
\overline{C}_{n,m}=\frac{1}{\int \overline{\gamma}(\overline{\mathbf{x}}_{n,m},1)\mathrm{d}\overline{\mathbf{x}}_{n,m}+\overline{\gamma}_{n,m}}.
\end{equation}

We summarize the belief propagation-based passive target tracking algorithm for a single time instant in Algorithm \ref{alg1}. As in \cite{leitinger2019belief}, a pruning operation is included in Step 15 to remove the PSs whose existence probability falls below a threshold $p_\mathrm{pru}$, preventing an unbounded increase in the number of PSs over time. Nevertheless, the algorithm remains difficult to implement in practice because the integrals involved in the messages cannot be computed directly.  In the following section, we will introduce a particle-based implementation of the algorithm to address this issue.

\begin{algorithm}[t]
\caption{Belief propagation-based passive target tracking at time $n$}
\label{alg1}
\begin{algorithmic}
\STATE \textbf{Input:} The beliefs $q(\mathbf{x}_{n-1,0})$ and $\underline{q}(\mathbf{x}_{n-1,k},r_{n-1,k})$ for $k\in\{1,\cdots,K_{n-1}\}$, and the measurement $\mathbf{z}_n$.\\
\STATE \textbf{Output:} The beliefs $q(\mathbf{x}_{n,0})$, $\underline{q}(\mathbf{x}_{n,k},r_{n,k})$ for all $k\in\{1,\cdots,K_n\}$, and the estimates of scatterer positions $\hat{\mathbf{x}}_{n,k}$ for all $k\in\{1,\cdots,K_n\}$.\\
\STATE \textbf{Step 1:}  Compute $\alpha(\mathbf{x}_{n,0})$ using (\ref{message alpha}).\\
\STATE \textbf{Step 2:}  Compute $\underline{\alpha}(\underline{\mathbf{x}}_{n,k},\underline{r}_{n,k})$ for $k\in\{1,\cdots,K_{n-1}\}$ using (\ref{predict PS})--(\ref{predict probability ps0}).
\STATE \textbf{Step 3:} Compute $\iota_k(\mathbf{x}_{n,0})$ for $k\in\{1,\cdots,K_{n-1}\}$ and $\kappa_m(\mathbf{x}_{n,0})$ for $m\in\{1,\cdots,M_n\}$ using (\ref{iota kappa}).
\STATE \textbf{Step 4:} Compute $\beta(a_{n,k})$ for $k\in\{1,\cdots,K_{n-1}\}$ using (\ref{message beta}).
\STATE \textbf{Step 5:} Compute $\xi(b_{n,m})$ for $m\in\{1,\cdots,M_n\}$ using (\ref{message xi})--(\ref{message xi 0}).
\STATE \textbf{Step 6:} Perform iterative association between legacy PSs and measurements using (\ref{iterative1})--(\ref{iterative2}).
\STATE \textbf{Step 7:} Compute $\eta(a_{n,k})$ for $k\in\{1,\cdots,K_{n-1}\}$ and $\varsigma(b_{n,m})$ for $m\in\{1,\cdots,M_n\}$ using (\ref{message eta}) and (\ref{message varsigma}), respectively.
\STATE \textbf{Step 8:} Compute $\gamma_k(\mathbf{x}_{n,0})$ for $k\in\{1,\cdots,K_{n-1}\}$ using (\ref{message gamma_k}).
\STATE \textbf{Step 9:} Compute $\underline{\gamma}(\underline{\mathbf{x}}_{n,k},\underline{r}_{n,k})$ for $k\in\{1,\cdots,K_{n-1}\}$ using (\ref{message legacy gamma 1})--(\ref{message legacy gamma 0}).
\STATE \textbf{Step 10:} Compute $\overline{\gamma}(\overline{\mathbf{x}}_{n,m},\overline{r}_{n,m})$ for $m\in\{1,\cdots,M_n\}$ using (\ref{message new phi 1})--(\ref{message new phi 0}).
\STATE \textbf{Step 11:} Compute $q(\mathbf{x}_{n,0})$ using (\ref{illuminator belief}).
\STATE \textbf{Step 12:} Compute $\underline{q}(\underline{\mathbf{x}}_{n,k},\underline{r}_{n,k})$ and $\underline{q}_{n,k}$ for $k\in\{1,\cdots,K_{n-1}\}$ using (\ref{legacy belief})--(\ref{normalization constant legacy}).
\STATE \textbf{Step 13:} Compute $\overline{q}(\overline{\mathbf{x}}_{n,m},\overline{r}_{n,m})$ and $\overline{q}_{n,m}$ for $m\in\{1,\cdots,M_n\}$ using (\ref{new belief})--(\ref{normalization constant new}).
\STATE \textbf{Step 14:} Let $K_n=K_{n-1}+M_n$. For $k=\{1,\cdots,K_{n-1}\}$, define $\underline{q}(\mathbf{x}_{n,k},r_{n,k})\triangleq\underline{q}(\underline{\mathbf{x}}_{n,k},\underline{r}_{n,k})\big|_{\underline{\mathbf{x}}_{n,k}=\mathbf{x}_{n,k},\underline{r}_{n,k}=r_{n,k}}$. For $m=\{1,\cdots,M_n\}$, define $\underline{q}(\mathbf{x}_{n,K_{n-1}+m},r_{n,K_{n-1}+m})\triangleq\overline{q}(\overline{\mathbf{x}}_{n,m},\overline{r}_{n,m})\big|_{\overline{\mathbf{x}}_{n,m}=\mathbf{x}_{n,K_{n-1}+m},\overline{r}_{n,m}=r_{n,K_{n-1}+m}}$, and $\underline{q}_{n,K_{n-1}+m}\triangleq\overline{q}_{n,m}$.
\STATE \textbf{Step 15:} For $k\in\{1,\cdots,K_n\}$, compute $p(r_{n,k}=1|\mathbf{z}_{1:n})$ as $1-\underline{q}_{n,k}$, and remove the $k$-th PS if $p(r_{n,k}=1|\mathbf{z}_{1:n})<p_\mathrm{pru}$.  Re-index the remaining PSs and accordingly update the total number of PSs to the new $K_n$.
\STATE \textbf{Step 16:} For $k\in\{1,\cdots,K_n\}$, determine whether the PS exists by evaluating if $p(r_{n,k}=1|\mathbf{z}_{1:n})>p_\mathrm{exi}$, and
compute $\hat{\mathbf{x}}_{n,k}$ for the existing PS using (\ref{scatterer estimate})--(\ref{conditional scatterer}) with $f(\mathbf{x}_{n,k},r_{n,k}=1|\mathbf{z}_{1:n})$ replaced by $\underline{q}(\mathbf{x}_{n,k},r_{n,k}=1)$.
    \end{algorithmic}
\end{algorithm}

\section{Particle-Based Implementation}\label{particle implementation}

We present a particle-based implementation to compute the messages derived in the previous section, primarily utilizing techniques introduced in \cite{meyer2017scalable,meyer2015distributed}, the supplementary material of \cite{venus2024graph}, and the publicly available code of \cite{leitinger2019belief}. In this approach, the belief of the transmitter position is represented by a set of discrete particles and their associated weights $\{(\mathbf{x}_{n,0}^{(s)},w_{n,0}^{(s)})\}_{s=1}^S$, where $S$ is the number of particles and the weights are normalized such that $\sum_{s=1}^S w_{n,0}^{(s)}=1$ to emulate the probability mass. Similarly, for each PS, the belief is represented by $\{(\mathbf{x}_{n,k}^{(s)},w_{n,k}^{(s)})\}_{s=1}^S$ with the weights satisfying $\sum_{s=1}^S w_{n,k}^{(s)}=1-\tilde{\underline{q}}_{n,k}$, where $\tilde{\underline{q}}_{n,k}$ is the particle-based approximation of $\underline{q}_{n,k}$. 
This equation implicitly reflects the existence probability of the PS. It is worth noting that a resampling step is performed to avoid particle degeneracy \cite{arulampalam2002tutorial} after obtaining the particle representation of each message. As a result, the particle weights become equal, 
and unless specified otherwise, we will not explicitly include them in the weight update equations when the resampled particles are subsequently used 
to compute 
other messages. In addition, with a slight abuse of notation, we will consistently use $\mathbf{x}_{n,0}^{(s)}$ to represent a particle for the transmitter position and $\mathbf{x}_{n,k}^{(s)}$ for the position of a PS in all relevant messages, although the resampling step changes the values of these particles.

\subsubsection{Prediction} Given the resampled particles $\{\mathbf{x}_{n-1,0}^{(s)}\}_{s=1}^S$ representing the belief of the transmitter position at time $n-1$, the particle $\mathbf{x}_{n,0}^{(s)}$ at time $n$ is sampled from the pdf $f(\mathbf{x}_{n,0}|\mathbf{x}_{n-1,0}^{(s)})$ for each $s\in\{1,\cdots,S\}$, with equal weights for all particles. This method is equivalent to choosing the transition pdf as the proposal distribution for a single time instant in the conventional sequential importance sampling particle filter \cite{arulampalam2002tutorial}. For each legacy PS, the particle $\underline{\mathbf{x}}_{n,k}^{(s)}$ is sampled from the transition pdf $f(\underline{\mathbf{x}}_{n,k}|\mathbf{x}_{n-1,k}^{(s)})$. To account for the survival probability, the weight is given by $w_{n,k}^{\underline{\alpha}(s)}=p_\mathrm{s} w_{n-1,k}^{(s)}=\frac{1}{S}p_\mathrm{s}(1-\tilde{\underline{q}}_{n-1,k})$. The particle-based approximation of the predicted non-existence probability in (\ref{predict probability ps0}) is calculated as $\tilde{\underline{\alpha}}_{n,k}=1-\sum_{s=1}^S w_{n,k}^{\underline{\alpha}(s)}$.

\subsubsection{Measurement evaluation} For the messages in (\ref{iota kappa}), we compute their weights using the importance sampling method with the proposal distribution chosen as $\alpha(\mathbf{x}_{n,0})$. Therefore, we have $w_{n,0}^{\iota_k(s)}\propto f_0(z_{n,0}|\mathbf{x}_{n,0}^{(s)})$ and $w_{n,0}^{\kappa_m(s)}\propto f_0(z_{n,0}|\mathbf{x}_{n,0}^{(s)})$ for all $k$ and $m$. The weights of each message are then normalized to sum to $1$. For the message $\beta(a_{n,k})$ in (\ref{message beta}), we use the technique introduced in \cite{meyer2015distributed} by stacking the resampled
particles of the positions of the transmitter and the legacy PS together as $\{(\mathbf{x}_{n,0}^{(s)},\underline{\mathbf{x}}_{n,k}^{(s)})\}_{s=1}^S$ with the weight of each stacked particle equal to $\underline{w}_{n,k}^{\prime(s)}=\frac{1}{S}(1-\tilde{\underline{\alpha}}_{n,k})$, which can be interpreted as the particle representation for the product of $\iota_k(\mathbf{x}_{n,0})$ and $\underline{\alpha}(\underline{\mathbf{x}}_{n,k},1)$. The particle-based approximation of $\beta(a_{n,k})$ can then be expressed as
\begin{equation}\label{beta tilde}
\begin{aligned}
\tilde{\beta}(a_{n,k})=\sum_{s=1}^Sg(\mathbf{x}_{n,0}^{(s)},\underline{\mathbf{x}}_{n,k}^{(s)},1,a_{n,k};\mathbf{z}_n)\underline{w}_{n,k}^{\prime(s)}+1(a_{n,k})\tilde{\underline{\alpha}}_{n,k}.
\end{aligned}
\end{equation}
To compute the message $\xi(b_{n,m})$ with $b_{n,m}=0$ in (\ref{message xi 0}), for each particle $\mathbf{x}_{n,0}^{(s)}$ representing the message $\kappa_m(\mathbf{x}_{n,0})$, we first sample a particle for the position of the new PS, denoted by $\overline{\mathbf{x}}_{n,m}^{(s)}$, from the pdf $f_\mathrm{n}(\overline{\mathbf{x}}_{n,m}|\mathbf{x}_{n,0}^{(s)})$, and then stack the particles as $\{(\mathbf{x}_{n,0}^{(s)},\overline{\mathbf{x}}_{n,m}^{(s)})\}_{s=1}^S$ with the weight of each stacked particle equal to $\overline{w}^{\prime(s)}_{n,m}=\frac{1}{S}$, which forms the particle representation for the product of $f_\mathrm{n}(\overline{\mathbf{x}}_{n,m}|\mathbf{x}_{n,0})$ and $\kappa_m(\mathbf{x}_{n,0})$. Then the message in (\ref{message xi 0}) can be approximated as
\begin{equation}\label{xi tilde}
\tilde{\xi}(b_{n,m})=1+\frac{\mu_{\mathrm{n},n}}{\mu_\mathrm{FA}f_\mathrm{FA}(\mathbf{z}_{n,m})}\sum_{s=1}^Sf(\mathbf{z}_{n,m}|\mathbf{x}_{n,0}^{(s)},\overline{\mathbf{x}}_{n,m}^{(s)})\overline{w}_{n,m}^{\prime(s)}.
\end{equation}

\subsubsection{Association between legacy PSs and measurements} The messages in (\ref{iterative1})--(\ref{message varsigma}) can be directly computed by replacing $\xi(b_{n,m})$ and $\beta(a_{n,k})$ with $\tilde{\xi}(b_{n,m})$ in (\ref{xi tilde}) and $\tilde{\beta}(a_{n,k})$ in (\ref{beta tilde}), respectively. The detailed steps will not be repeated here.

\subsubsection{Measurement update and belief calculation} Substituting (\ref{message gamma_k}) into (\ref{illuminator belief}) and performing some rearrangement, we obtain
\begin{equation}\label{arrange}
\begin{aligned}
&q(\mathbf{x}_{n,0})\propto\int\cdots\int\alpha(\mathbf{x}_{n,0})f_0(z_{n,0}|\mathbf{x}_{n,0})\\
&~~~~~~~~~~~~~~~~~~~~~~\times\prod_{k=1}^{K_{n-1}}\Bigg((1-\underline{\alpha}_{n,k})\sum_{a_{n,k}=0}^{M_n}\eta(a_{n,k})\\
&~~~~~~~~~~~~~~~~~~~~~~\times g(\mathbf{x}_{n,0},\underline{\mathbf{x}}_{n,k},1,a_{n,k};\mathbf{z}_n)+\eta(0) \underline{\alpha}_{n,k}\Bigg)\\
&~~~~~~~~~~~~~~~~~~~~~~\times\underline{\alpha}(\underline{\mathbf{x}}_{n,k},1)\mathrm{d}\underline{\mathbf{x}}_{n,1}\cdots\mathrm{d}\underline{\mathbf{x}}_{n,K_{n-1}}.
\end{aligned}
\end{equation}
In the above equation, the particles representing $\alpha(\mathbf{x}_{n,0})f_0(z_{n,0}|\mathbf{x}_{n,0})$ can be generated in the same manner as those representing $\iota_k(\mathbf{x}_{n,0})$ or $\kappa_m(\mathbf{x}_{n,0})$. Next, we stack the particle $\mathbf{x}_{n,0}^{(s)}$ representing $\alpha(\mathbf{x}_{n,0})f_0(z_{n,0}|\mathbf{x}_{n,0})$ and the particles $\underline{\mathbf{x}}_{n,k}^{(s)}$ representing $\underline{\alpha}(\underline{\mathbf{x}}_{n,k},1)$ for all $k\in\{1,\cdots,K_{n-1}\}$ as $\{(\mathbf{x}_{n,0}^{(s)},\underline{\mathbf{x}}_{n,1}^{(s)},\cdots,\underline{\mathbf{x}}_{n,K_{n-1}}^{(s)})\}_{s=1}^S$, which forms the particle representation of $\alpha(\mathbf{x}_{n,0})f_0(z_{n,0}|\mathbf{x}_{n,0})\prod_{k=1}^{K_{n-1}}\underline{\alpha}(\underline{\mathbf{x}}_{n,k},1)$. The weights of the stacked particles are all equal because of the resampling step performed for each message. Since the belief $q(\mathbf{x}_{n,0})$ will eventually be normalized, the exact values of these weights are not of concern. 
We then apply importance sampling to generate the particle representation of the expression inside the integral in (\ref{arrange}), with the weight given by
\begin{equation}\label{stack weight}
\begin{aligned}
&w_{n,0}^{(s)}\propto \prod_{k=1}^{K_{n-1}}\Bigg((1-\tilde{\underline{\alpha}}_{n,k})\sum_{a_{n,k}=0}^{M_n}\tilde{\eta}(a_{n,k})\\
&~~~~~~~~~~~~~~~~~~\times g(\mathbf{x}_{n,0}^{(s)},\underline{\mathbf{x}}_{n,k}^{(s)},1,a_{n,k};\mathbf{z}_n)+\tilde{\eta}(0)\tilde{\underline{\alpha}}_{n,k}\Bigg),
\end{aligned}
\end{equation}
where $\tilde{\eta}(a_{n,k})$ is the particle-based approximation of $\eta(a_{n,k})$. Finally, the particles representing $q(\mathbf{x}_{n,0})$ can be obtained by removing $\{\underline{\mathbf{x}}_{n,k}^{(s)}\}_{k=1}^{K_{n-1}}$ from the stacked particles while keeping the weight in (\ref{stack weight}) unchanged, which serves as the Monte Carlo implementation for the marginalization in (\ref{arrange}). For the belief of the augmented state of the legacy PS, we substitute (\ref{message legacy gamma 1}) into (\ref{legacy belief}) and obtain
\begin{equation}\label{legacy belief substitution}
\begin{aligned}
\underline{q}(\underline{\mathbf{x}}_{n,k},1)\propto &\int\sum_{a_{n,k}=0}^{M_n}\eta(a_{n,k})g(\mathbf{x}_{n,0},\underline{\mathbf{x}}_{n,k},1,a_{n,k};\mathbf{z}_n)\\
&~~~~~\times\iota_k(\mathbf{x}_{n,0})\underline{\alpha}(\underline{\mathbf{x}}_{n,k},1)\mathrm{d}\mathbf{x}_{n,0}.
\end{aligned}
\end{equation}
We use $\iota_k(\mathbf{x}_{n,0})\underline{\alpha}(\underline{\mathbf{x}}_{n,k},1)$ as the proposal distribution, whose particle representation was obtained during the calculation of $\tilde{\beta}(a_{n,k})$, to generate particles for the expression inside the integral in (\ref{legacy belief substitution}), and the weight is given by
\begin{equation}
\underline{w}_{n,k}^{\prime\prime (s)}=\underline{w}_{n,k}^{\prime(s)}\sum_{a_{n,k}=0}^{M_n}\tilde{\eta}(a_{n,k})g(\mathbf{x}_{n,0}^{(s)},\underline{\mathbf{x}}_{n,k}^{(s)},1,a_{n,k};\mathbf{z}_n).
\end{equation}
Next, we perform marginalization by removing $\underline{\mathbf{x}}_{n,k}^{(s)}$ from the stacked particles to obtain the particle representation of the right-hand side of (\ref{legacy belief substitution}). 
The weights are then normalized as $\underline{w}_{n,k}^{(s)}=\tilde{\underline{C}}_{n,k}\underline{w}_{n,k}^{\prime\prime(s)}$, where
\begin{equation}
\underline{\tilde{C}}_{n,k}=\frac{1}{\sum_{s=1}^S\underline{w}_{n,k}^{\prime\prime(s)}+\tilde{\eta}(0)\tilde{\underline{\alpha}}_{n,k}}
\end{equation}
is the approximated normalization constant in (\ref{normalization constant legacy}).
For the belief of the augmented state of the new PS, we combine (\ref{h1}), (\ref{message new phi 1}), and (\ref{new belief}) to obtain
\begin{equation}\label{new belief substitution}
\begin{aligned}
&\overline{q}(\overline{\mathbf{x}}_{n,m},1)\propto\int \frac{\varsigma(0)\mu_{\mathrm{n},n}}{\mu_{\mathrm{FA}}f_{\mathrm{FA}}(\mathbf{z}_{n,m})}f(\mathbf{z}_{n,m}|\mathbf{x}_{n,0},\overline{\mathbf{x}}_{n,m})\\
&~~~~~~~~~~~~~~~~~~~\times f_\mathrm{n}(\overline{\mathbf{x}}_{n,m}|\mathbf{x}_{n,0})\kappa_m(\mathbf{x}_{n,0})\mathrm{d}\mathbf{x}_{n,0}.
\end{aligned}
\end{equation}
The particle representation for the expression inside the integral in (\ref{new belief substitution}) can be obtained via importance sampling with the proposal distribution chosen as $f_\mathrm{n}(\overline{\mathbf{x}}_{n,m}|\mathbf{x}_{n,0})\kappa_m(\mathbf{x}_{n,0})$,
and the weight is given by
\begin{equation}
\overline{w}^{\prime\prime(s)}_{n,m}=\overline{w}^{\prime(s)}_{n,m}\frac{\tilde{\varsigma}(0)\mu_{\mathrm{n},n}}{\mu_{\mathrm{FA}}f_{\mathrm{FA}}(\mathbf{z}_{n,m})}f(\mathbf{z}_{n,m}|\mathbf{x}_{n,0}^{(s)},\overline{\mathbf{x}}_{n,m}^{(s)}),
\end{equation}
where we recall that $\overline{w}^{\prime (s)}_{n,m}$ was derived during the calculation of $\tilde{\xi}(b_{n,m})$.
We again perform marginalization by removing $\overline{\mathbf{x}}_{n,m}^{(s)}$ from the stacked particles while keeping the weights unchanged. The weights are then normalized as $\overline{w}_{n,m}^{(s)}=\tilde{\overline{C}}_{n,m}\overline{w}_{n,m}^{\prime\prime(s)}$, with the normalization constant in (\ref{normalization constant new}) approximated as
\begin{equation}
\tilde{\overline{C}}_{n,m}=\frac{1}{\sum_{s=1}^S\overline{w}_{n,m}^{\prime\prime(s)}+\sum_{b_{n,m}=0}^{K_{n-1}}\tilde{\varsigma}(b_{n,m})},
\end{equation}
where we have used the relation $\overline{\gamma}_{n,m}=\sum_{b_{n,m}=0}^{K_{n-1}}\varsigma(b_{n,m})$.

Once the beliefs for the positions of the transmitter and PSs at time $n$ are obtained, their estimates can be readily computed by taking the arithmetic mean of the resampled particles. The particles for the legacy PSs $\{\underline{\mathbf{x}}_{n,k}^{(s)},\underline{w}_{n,k}^{(s)}\}_{s=1}^S$ for $k\in\{1,\cdots,K_{n-1}\}$ and the new PSs $\{\overline{\mathbf{x}}_{n,m}^{(s)},\overline{w}_{n,m}^{(s)}\}_{s=1}^S$ for $m\in\{1,\cdots,M_n\}$ are then combined to form the particles for the legacy PSs $\{\mathbf{x}_{n,k}^{(s)},w_{n,k}^{(s)}\}_{s=1}^S$ at time $n+1$ with $k\in\{1,\cdots,K_n\}$.
It is worth noting that the computational complexity of the whole particle-based implementation scales linearly with the number of particles, due to the employed stacking technique, which enhances the scalability of the algorithm.

\section{Simulations}\label{simulation}

\begin{figure}
\centering
\includegraphics[width=1\columnwidth]{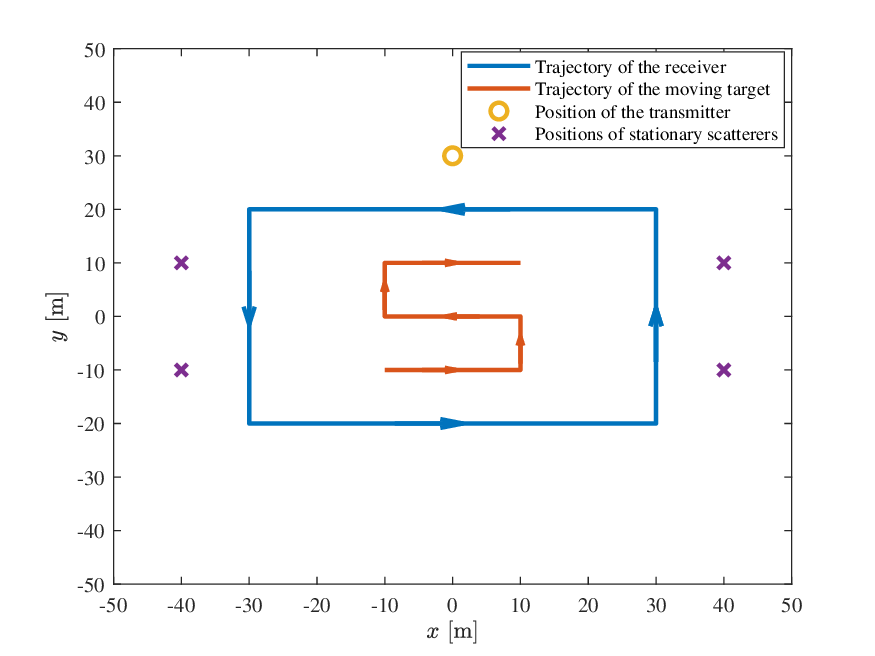}
\caption{Illustration of the passive target tracking scenario.}
\label{model illustration}
\end{figure}

 In this section, we evaluate the performance of the proposed passive target tracking algorithm through simulations. In the following, the unit for all distance-related parameters is meters. We consider a scenario as shown in Fig. \ref{model illustration}, where the transmitter is located at $[0,30]^\mathrm{T}$, and four stationary scatterers are located at $[40,10]^\mathrm{T}$, $[40,-10]^\mathrm{T}$, $[-40,-10]^\mathrm{T}$, and $[-40,10]^\mathrm{T}$. The moving target starts at $[-10,-10]^\mathrm{T}$ and follows a path through $[10,-10]^\mathrm{T}$, $[10,0]^\mathrm{T}$, $[-10,0]^\mathrm{T}$, $[-10,10]^\mathrm{T}$, and finally reaches $[10,10]^\mathrm{T}$. 
 Meanwhile, the mobile receiver starts at $[0,-20]^\mathrm{T}$ and moves through $[30,-20]^\mathrm{T}$, $[30,20]^\mathrm{T}$, $[-30,20]^\mathrm{T}$, $[-30,-20]^\mathrm{T}$, and then back to its starting position at $[0, -20]^\mathrm{T}$. The simulation runs for $200$ time instants, with the target moving $0.4$ meters and the receiver moving $1$ meter per time instant. In each time instant, the ground truth of the relative distance and AOA parameters is generated according to (\ref{relative distance}) and (\ref{AOA}), respectively. The AOA measurement of the direct path is given by $z_{n,0}=\theta_{n,0}+n_{\theta,n,0}$, where $n_{\theta,n,0}$ is white Gaussian noise with zero mean and variance $\sigma_\theta^2$. For a scattered path, the relative distance and AOA measurements are given by $z_{d,n,m}=d_{n,m}+n_{d,n,m}$ and $z_{\theta,n,m}=\theta_{n,m}+n_{\theta,n,m}$, with $n_{d,n,m}$ and $n_{\theta,n,m}$ representing white Gaussian noise with zero mean and variances $\sigma_d^2$ and $\sigma_\theta^2$, respectively. In the simulations, we choose $\sigma_d=0.1$ and $\sigma_\theta=\frac{\pi}{180}$ for generating these measurements. The mean parameter of the Poisson distribution for the number of false alarm measurements is set to $\mu_\mathrm{FA}=1$. For each false alarm measurement, the relative distance is uniformly distributed between $0$ and $50$, and the AOA is uniformly distributed between $0$ and $\pi$. The detection probability of the true scattered paths is set to a constant value of $0.95$.

 In the message passing algorithm, the parameters are set identically to those used for generating the measurements, expect that $\sigma_d$ is adjusted to $0.2$ and $\sigma_\theta$ to $\frac{\pi}{90}$ in the likelihood function to better accommodate bad measurement samples. Additionally, the survival probability is set to $p_\mathrm{s}=0.999$. The transition pdf of the transmitter position is chosen as $f(\mathbf{x}_{n,0}|\mathbf{x}_{n-1,0})=\mathcal{N}(\mathbf{x}_{n,0};\mathbf{x}_{n-1,0},\sigma_0^2\mathbf{I}_2)$ with $\sigma_0=0.1$, and the transition pdf of the position of a legacy PS is chosen as $f(\underline{\mathbf{x}}_{n,k}|\mathbf{x}_{n-1,k})=\mathcal{N}(\underline{\mathbf{x}}_{n,k};\mathbf{x}_{n-1,k},\sigma_k^2\mathbf{I}_2)$ with $\sigma_k=0.5$ for all $k$. 
 As in \cite{leitinger2019belief}, we adopt a probability hypothesis density (PHD) filter \cite{vo2003sequential} to determine the mean parameter $\mu_{\mathrm{n},n}$ of the Poisson distribution for the number of measurements generated by new PSs, in which the intensity functions for both undetected PSs and newly born PSs are assumed uniform over the area $\{[x,y]^\mathrm{T}|-50\leq x\leq50,-50\leq y\leq 50\}$, with the initial values of the mean parameters for the numbers of undetected PSs and newly born PSs set to $5$ and $1\times10^{-4}$, respectively. For the pdf $f_\mathrm{n}(\overline{\mathbf{x}}_{n,m}|\mathbf{x}_{n,0})$, instead of assuming a uniform distribution as in the PHD filter, we determine it through geometric principles based on the relative distance and AOA measurements, which will be specified later in the particle form. The iterative association between legacy PSs and measurements is performed using the code released by the authors of \cite{leitinger2019belief}, with the stopping criterion identical to that used in their implementation. In each time instant, we remove the legacy PSs with posterior existence probability smaller than $p_{\mathrm{pru}}<1\times10^{-3}$.
 
 The number of particles is set to $S=1000$. Given $\mathbf{x}_{n,0}^{(s)}$, when drawing a particle from $f_\mathrm{n}(\overline{\mathbf{x}}_{n,m}|\mathbf{x}_{n,0}^{(s)})$, we first introduce noise to the measured relative distance and AOA as $z_{d,n,m}^{(s)}=z_{d,n,m}+n^{(s)}_{d,n,m}$ and $z_{\theta,n,m}^{(s)}=z_{\theta,n,m}+n^{(s)}_{\theta,n,m}$, where $n^{(s)}_{d,n,m}$ and $n^{(s)}_{\theta,n,m}$ are Gaussian noise with zero mean and variance $\sigma_d^2$ and $\sigma_\theta^2$, respectively. Each particle $\overline{\mathbf{x}}_{n,m}^{(s)}$ representing the position of the new PS is then calculated using geometric relationships with $z_{d,n,m}^{(s)}$ and $z_{\theta,n,m}^{(s)}$ treated as the true relative distance and AOA, respectively. Since it is unclear from which side a path arrives at the receiver's antenna array, we sample half of the particles on each side to account for this ambiguity. For initializing the particles of the transmitter position, we similarly add noise to the AOA of the direct path, and assume the propagation distance of the direct path to be uniformly distributed between $0$ to $150$. The transmitter position is then determined using geometric relationships, with half of the particles sampled on each side of the receiver's antenna array. Due to the large uncertainty in positions of both the transmitter and the PSs at the beginning, the belief propagation framework for their joint estimation fails to work effectively. Specifically, an inaccurate estimate of the transmitter position can adversely affect the estimation of the PS positions, and vice versa. Therefore, we initially focus solely on estimating the transmitter position based on the AOA measurement of the direct path. This estimation is performed using the transition model $f(\mathbf{x}_{n,0}|\mathbf{x}_{n-1,0})$ and the measurement model $f_0(z_{n,0}|\mathbf{x}_{n,0})$ via a conventional sequential importance sampling particle filter \cite{arulampalam2002tutorial}. Once the standard deviation of the transmitter position, given by $\sqrt{\mathrm{tr}\left(\frac{1}{S}\sum_{s=1}^S(\mathbf{x}_{n,0}^{(s)})(\mathbf{x}_{n,0}^{(s)})^\mathrm{T}\right)}$, falls below $5$, we proceed with the estimation of the PS positions. At this new stage, we will further explore two simplified variants of the belief propagation-based algorithm. The first method fixes the initially estimated transmitter position and updates only the beliefs of the PS positions. The second method updates the beliefs of both the transmitter position and the PS positions, but the transmitter position continues to be estimated solely based on the AOA measurement of the direct path. These two methods will be referred to as ``simplified 1'' and ``simplified 2,'' respectively, in the simulation results.

 We consider three additional benchmark schemes for comparison, including the Rao-Blackwellized (RB) particle filter in \cite{gentner2016multipath}, the first-order extended Kalman filter (EKF) in \cite{caceres2010hybrid}, and the geometry-based method in \cite{qian2018widar2}. As these methods lack inherent mechanisms to associate scatterers with measurements, we devise the following association strategies: For all benchmarks, we assume the true number of scatterers $L_n$ is known. For the RB particle filter, if the number of measurements of scattered paths $M_n$ is larger than $L_n$, we select $L_n$ from the $M_n$ measurements and perform the association by maximizing the joint likelihood function for all particles; otherwise, we select $M_n$ from the $L_n$ scatterers and perform the association in a similar manner. In the latter case, for the remaining $L_n-M_n$ scatterers without associated measurements, we skip the update step and retain the prior distribution from the prediction step as the posterior distribution of the scatterer positions, while the transmitter position is updated solely based on the scatterers with associated measurements. Additionally, to handle the large number of particles in the association process, we implement the Hungarian algorithm \cite{kuhn1955hungarian} to reduce computational complexity. 
 For the EKF, the association is also based on maximum likelihood, in which the true positions of scatterers are substituted. For the geometry-based method, the association is performed by minimizing the weighted sum of the relative changes in multipath parameters between the current and previous time instants, with weights set to the default values provided in the publicly available code of \cite{qian2018widar2}. We note that these association strategies do not account for false alarms and missed detections in the measurements, which leads to reduced effectiveness. 
 At the beginning stage of all three benchmarks, the transmitter position is estimated using the AOA measurements of the direct path, as in our proposed algorithm. Subsequently, the RB particle filter continues this estimation by integrating the measurement model of the direct path with the predicted pdf of the transmitter position to form a composite prediction step. In contrast, the EKF and the geometry-based method fix the transmitter position as a constant value. In addition, for these two methods, we initialize the scatterer positions without the ambiguity regarding the arrival direction relative to the receiver's antenna array. In the geometry-based method, the ambiguity is resolved in the subsequent tracking process by selecting scatterer positions that are closer to those estimated in the previous time instants, while the EKF inherently avoids the ambiguity during tracking. 

\begin{figure*}
	\normalsize
	\centering
	\subfigure[$n=1$]{
		\includegraphics[width=2.2in]{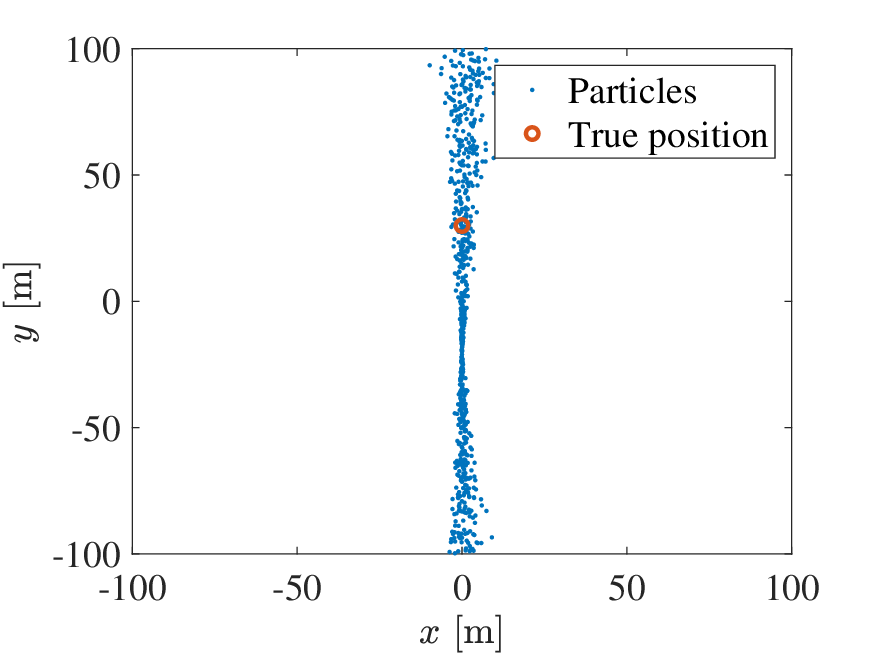}
	}
	\subfigure[$n=20$]{
		\includegraphics[width=2.2in]{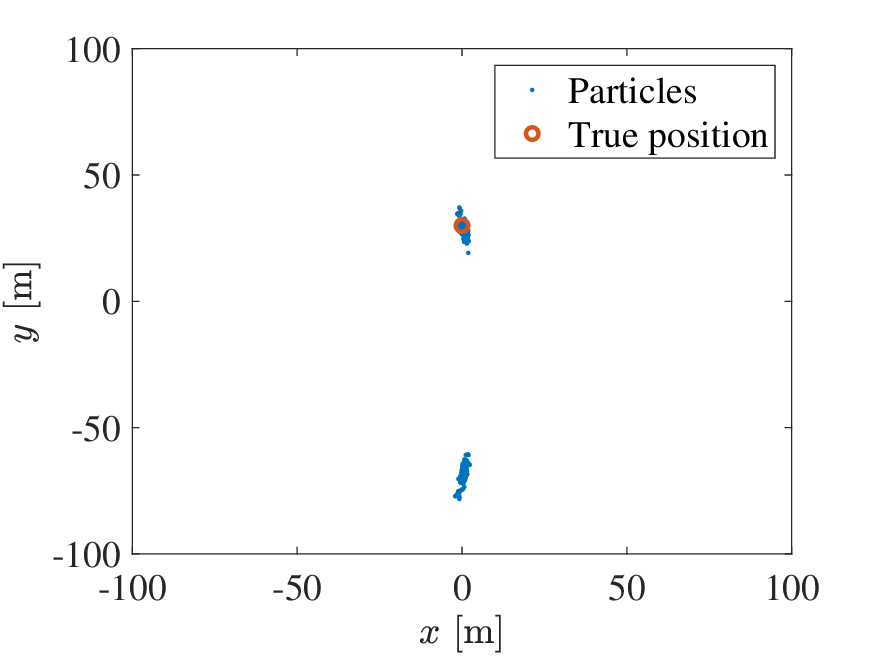}
	}
	\subfigure[$n=50$]{
		\includegraphics[width=2.2in]{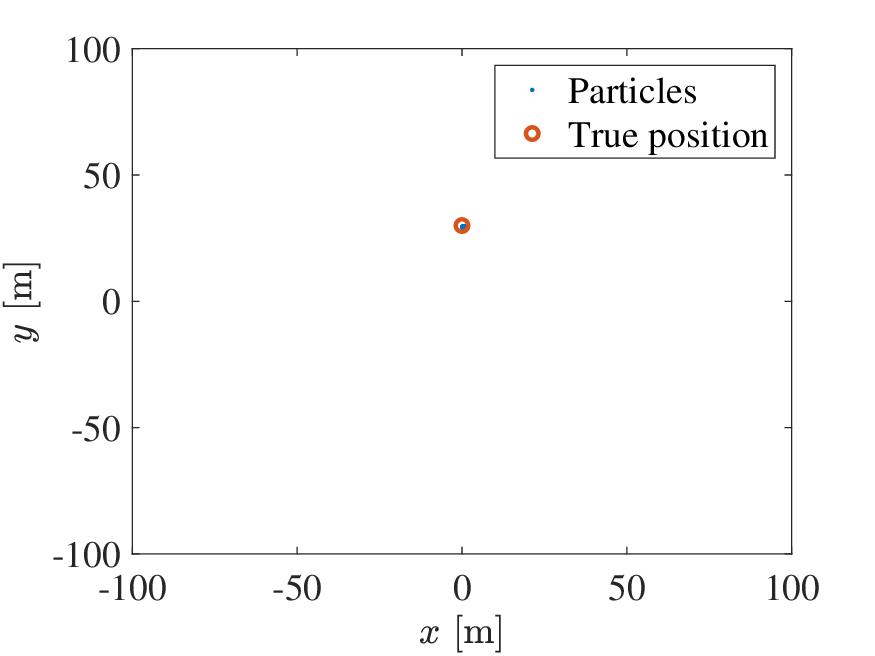}
	}
	\caption{Illustration of particles of the transmitter position at different time instants.}
	\label{illuminator particles}
\end{figure*}

\begin{figure*}
	\normalsize
	\centering
	\subfigure[$n=33$]{
		\includegraphics[width=2.2in]{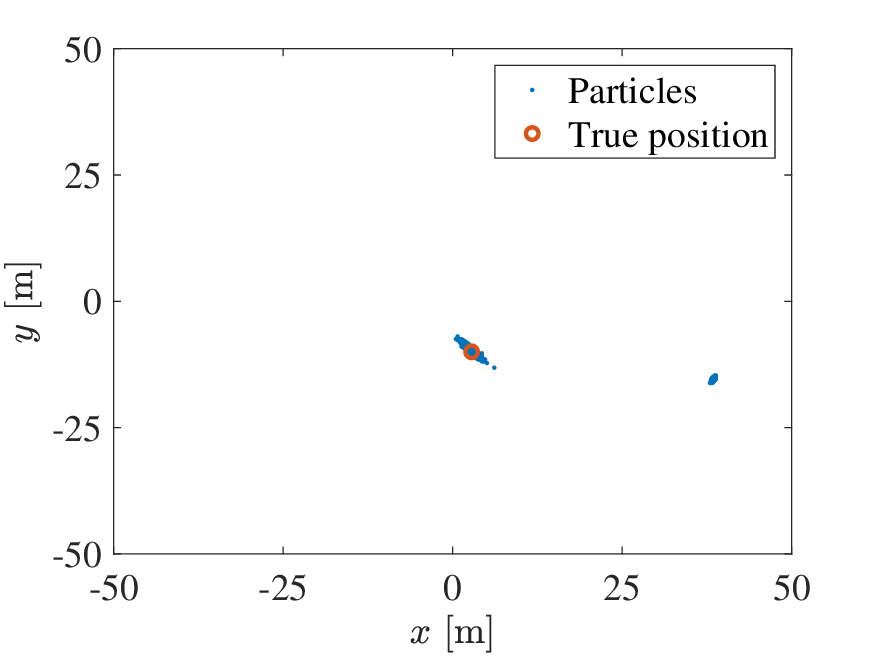}
	}
	\subfigure[$n=50$]{
		\includegraphics[width=2.2in]{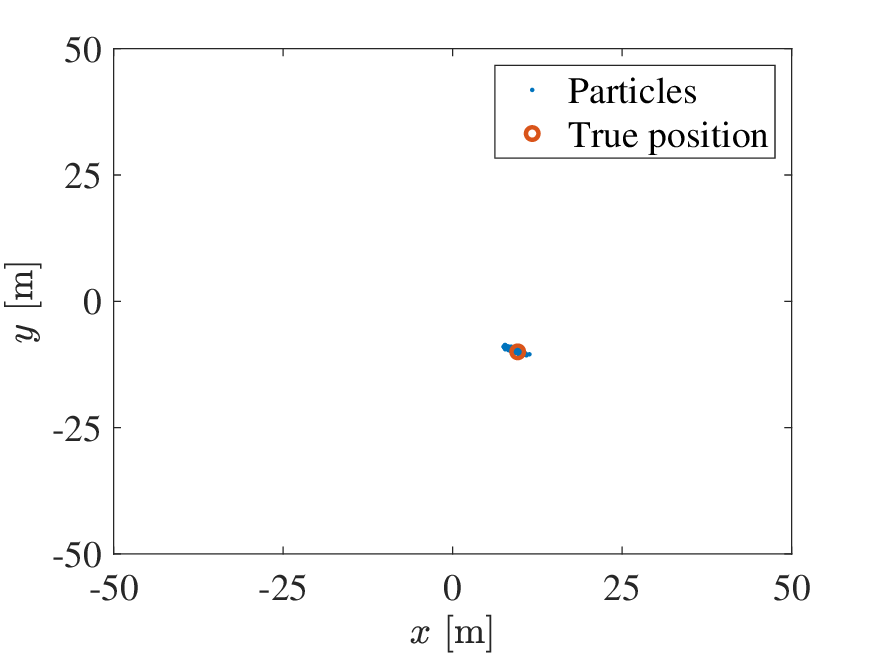}
	}
	\subfigure[$n=100$]{
		\includegraphics[width=2.2in]{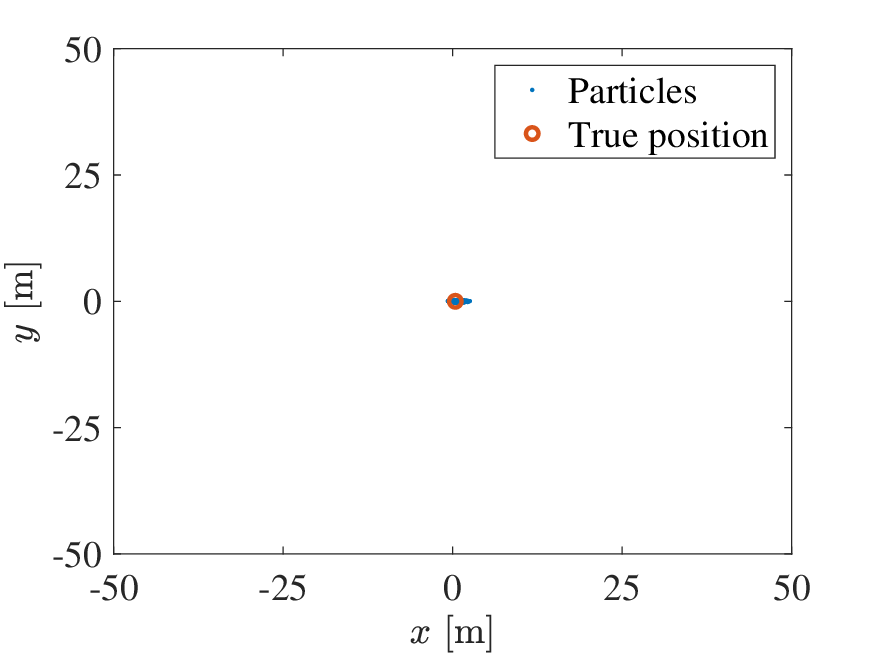}
	}
	\caption{Illustration of particles of the target position at different time instants.}
	\label{target particles}
	\hrulefill
\end{figure*}

 In Fig. \ref{illuminator particles}, we illustrate the particle distributions of the transmitter position at different time instants. At $n=1$, the particles are generated on both sides of the receiver's antenna array based on the AOA measurement of the direct path, and the distance to the receiver is assumed to follow a uniform distribution. This creates a large spread of particles, leading to significant uncertainty about the transmitter position. By time $n=20$, the distance uncertainty has been reduced, and the particles center around two possible positions due to the persisting AOA ambiguity---the receiver does not change its moving direction, thus offering no additional directional information. By time $n=50$, after the receiver changes its direction, the AOA uncertainty has been removed, and the particles converge around the true position of the transmitter.  In Fig. \ref{target particles}, we show the particle distributions of the target position at different time instants. The target tracking process begins at $n=33$, as the standard deviation of the transmitter position has reduced below the threshold of $5$. At this point, similar to Fig. \ref{illuminator particles}, the AOA ambiguity results in two possible positions for the target. As the receiver moves, this ambiguity can be resolved, allowing the particles to accurately converge around the true target positions at $n=50$ and $n=100$.

\begin{figure}
\centering
\includegraphics[width=1\columnwidth]{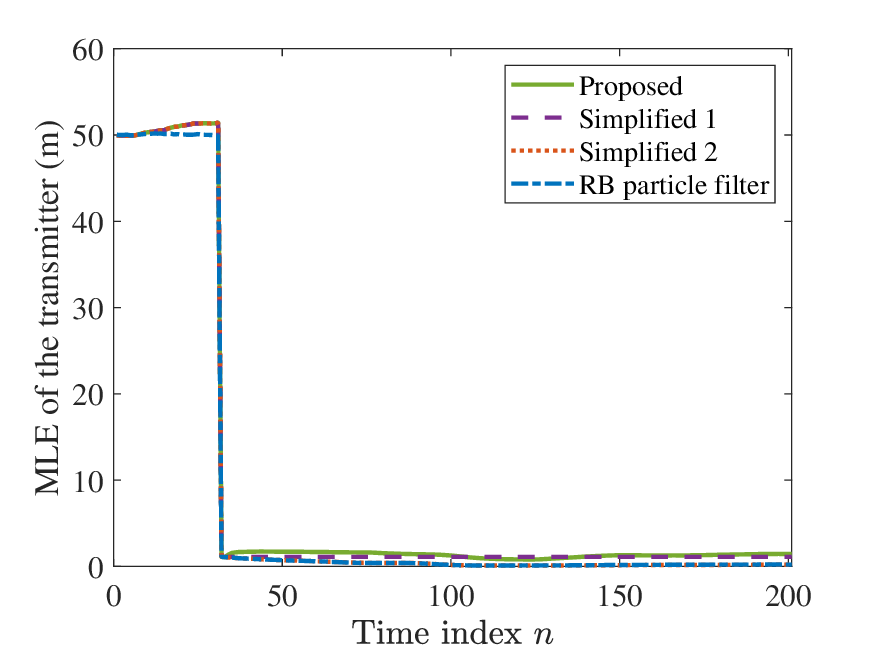}
\caption{Comparison of the MLE of the transmitter.}
\label{illuminator position error}
\end{figure}

In Fig. \ref{illuminator position error}, we plot the mean localization error (MLE) of the transmitter for different schemes, which is calculated as the average value of $\lVert\mathbf{x}_{n,0}-\hat{\mathbf{x}}_{n,0}\rVert_2$ over $1000$ simulations. It is worth mentioning that the first-order EKF and the geometry-based method use the same approach as simplified method 1 for estimating the transmitter position, and hence no separate curve is shown for these two methods. As illustrated, the error is large at the early stage due to the ambiguity from the AOA measurements. Once the receiver changes direction at $n=32$, the ambiguity is quickly resolved, leading to a significant reduction in localization error. In the following time instants, the RB particle filter and simplified method 2 show nearly identical performance, slightly outperforming the proposed method and simplified method 1. The RB particle filter performs well because a precise initial estimate is obtained based on the AOA of the direct path, and the subsequent adjustments are minor. In contrast, the ability of the proposed method to automatically identify the false alarms and missed detections leads to a slight compromise in the estimation accuracy of the transmitter position. However, as we will demonstrate later, the proposed method achieves superior target tracking performance compared with the RB particle filter.
 
\begin{figure}
\centering
\includegraphics[width=1\columnwidth]{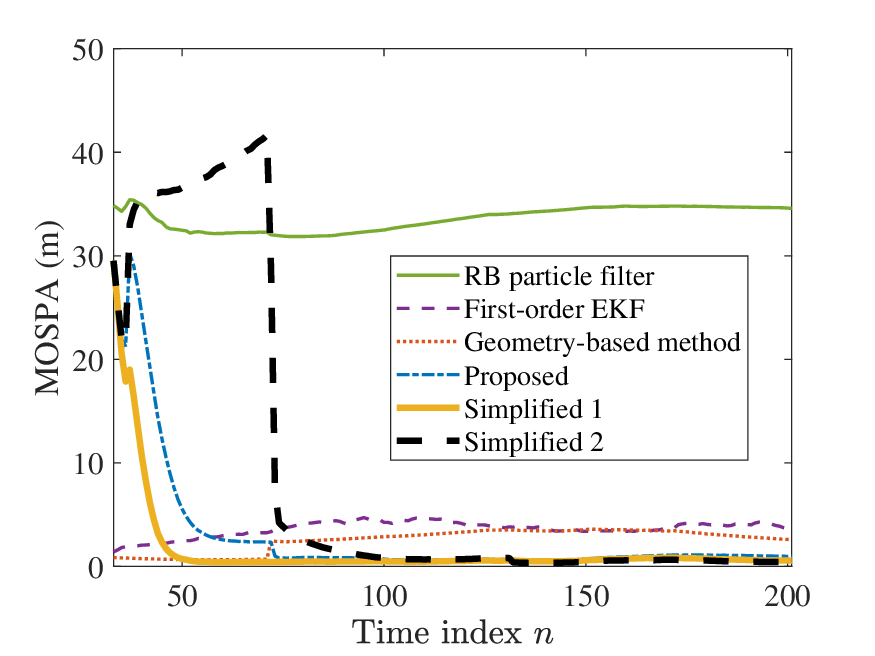}
\caption{Comparison of the MOSPA of all scatterers.}
\label{MOSPA}
\end{figure}

\begin{figure}
\centering
\includegraphics[width=1\columnwidth]{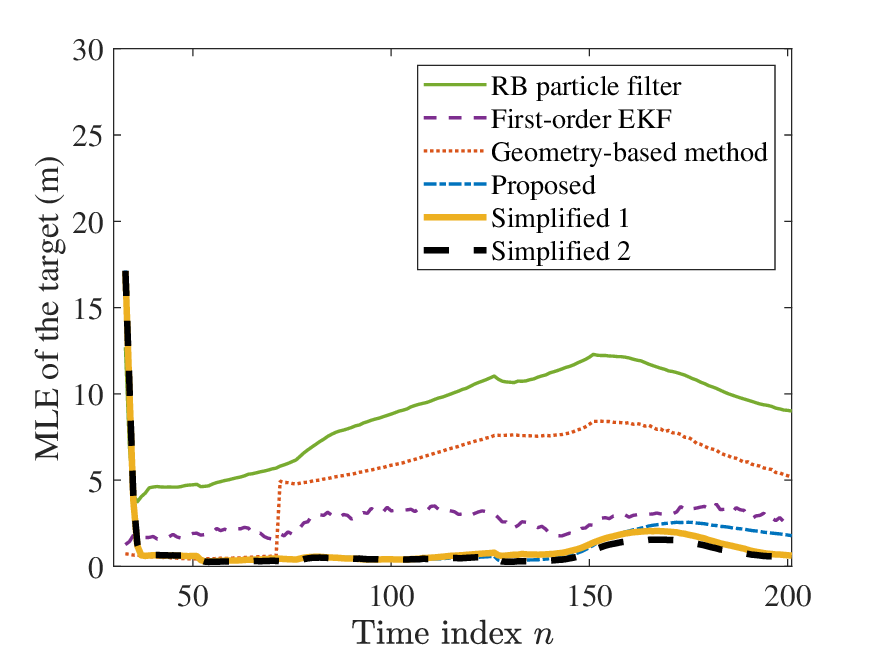}
\caption{Comparison of the MLE of the target.}
\label{target position error}
\end{figure}
In Fig. \ref{MOSPA} and Fig. \ref{target position error}, we compare the mean optimal subpattern assignment (MOSPA) error \cite{schuhmacher2008consistent} for all scatterers and the MLE of the target, respectively, across all schemes, with both metrics averaged over $1000$ simulations. For MOSPA, we set the order to $1$ and the cut-off parameter to $10$.  
We plot the figures starting from $n=33$, marking the beginning of the target tracking process. Since the three benchmarks are unable to identify false alarms and missed detections of measurements, their performance is inferior to the proposed method and its two simplified variants. The error curves do not decrease monotonically over time, as the localization accuracy is affected by the relative positions of the scatterers, the receiver, and the transmitter. The geometry-based method is particularly sensitive to these relative positions, and consequently, incorrect associations occur when the receiver changes its moving direction, such as at $n=72$, resulting in large errors. In Fig. \ref{MOSPA}, the proposed method and its two variants exhibit different convergence speeds. Our simulation results suggest that a smaller variance in the estimate of the transmitter position contributes to faster convergence of scatterer position estimates. Simplified method 1, which uses a fixed transmitter position, achieves the fastest convergence due to the zero variance. The proposed method has a smaller variance than simplified method 2, as the measurement update step in message passing helps reduce uncertainty in the estimates.

\section{Conclusions}\label{conclusion}
This paper presented a passive target tracking algorithm using RF signals from a transmitter with an unknown position. A mobile receiver collects signals emitted from the transmitter and scattered by the moving target and the stationary objects in the environment, and uses relative distance and AOA parameters extracted from these signals to jointly localize the transmitter and the scatterers over time. We established a probabilistic framework to fully characterize the target tracking model, incorporating the localization of the transmitter and scatterers, association between scatterers and multipath measurements, and the identification of false alarms and missed detections. We used a belief propagation-based message passing algorithm to compute the posterior distributions of the positions of scatterers and the transmitter, and then introduced a particle implementation for this algorithm. Simulation results have shown that our proposed method achieves superior performance compared to existing benchmarks.

\bibliography{document}
\bibliographystyle{IEEEtran}
\end{document}